\documentclass[aip,reprint,author-year,nofootinbib]{revtex4-1}

\usepackage{graphicx}
\usepackage{amsmath,amssymb}             
\usepackage{stmaryrd}					
\usepackage{color}
\usepackage[dvipsnames]{xcolor}
\usepackage{color,hyperref}
\definecolor{darkblue}{rgb}{0.0,0.0,0.4}
\definecolor{red}{rgb}{0.7,0.0,0.0}
\definecolor{green}{rgb}{0.0,0.5,0.0}
\hypersetup{colorlinks,breaklinks,
            linkcolor=darkblue,urlcolor=darkblue,
            anchorcolor=darkblue,citecolor=RoyalBlue}

\def\apjl{Astrophys. J. Lett.}
\def\apj{Astrophys. J.}

\def\apjs{Astrophys. J. Supp.}

\def\mnras{Mon. Not. R. Astron. Soc.}

\def\prd{Phys. Rev. D}
\def\prl{Phys. Rev. Lett.}
\def\physrep{Physics Reports}
\def\nat{Nature}

\def\aap{Astron. \& Astrophys.}

\def\jcap{Journal of Cosmology and Astroparticle Physics}
\def\na{New Astronomy}

\defcitealias{Jasche2015BORGSDSS}{\textsc{borg sdss}}
\newcommand{\borg}{\textsc{borg}}
\newcommand{\cola}{\textsc{cola}}

\begin{document}


\title{Bayesian analysis of the dynamic cosmic web in the SDSS galaxy survey}


\author{Florent Leclercq}
\email{florent.leclercq@polytechnique.org}
\affiliation{Institut d'Astrophysique de Paris (IAP), UMR 7095, CNRS -- UPMC Universit\'e Paris 6, Sorbonne Universit\'es, 98bis boulevard Arago, F-75014 Paris, France}
\affiliation{Institut Lagrange de Paris (ILP), Sorbonne Universit\'es,\\ 98bis boulevard Arago, F-75014 Paris, France}
\affiliation{\'Ecole polytechnique ParisTech,\\ Route de Saclay, F-91128 Palaiseau, France}

\author{Jens Jasche}
\affiliation{Excellence Cluster Universe, Technische Universit\"at M\"unchen,\\ Boltzmannstrasse 2, D-85748 Garching, Germany}

\author{Benjamin Wandelt}
\affiliation{Institut d'Astrophysique de Paris (IAP), UMR 7095, CNRS -- UPMC Universit\'e Paris 6, Sorbonne Universit\'es, 98bis boulevard Arago, F-75014 Paris, France}
\affiliation{Institut Lagrange de Paris (ILP), Sorbonne Universit\'es,\\ 98bis boulevard Arago, F-75014 Paris, France}
\affiliation{Department of Physics, University of Illinois at Urbana-Champaign,\\ 1110 West Green Street, Urbana, IL~61801, USA}
\affiliation{Department of Astronomy, University of Illinois at Urbana-Champaign,\\ 1002 West Green Street, Urbana, IL~61801, USA}


\date{\today}

\begin{abstract}
\noindent Recent application of the Bayesian algorithm {\borg} to the Sloan Digital Sky Survey (SDSS) main sample galaxies resulted in the physical inference of the formation history of the observed large-scale structure from its origin to the present epoch. In this work, we use these inferences as inputs for a detailed probabilistic cosmic web-type analysis. To do so, we generate a large set of data-constrained realizations of the large-scale structure using a fast, fully non-linear gravitational model. We then perform a dynamic classification of the cosmic web into four distinct components (voids, sheets, filaments, and clusters) on the basis of the tidal field. Our inference framework automatically and self-consistently propagates typical observational uncertainties to web-type classification. As a result, this study produces accurate cosmographic classification of large-scale structure elements in the SDSS volume. By also providing the history of these structure maps, the approach allows an analysis of the origin and growth of the early traces of the cosmic web present in the initial density field and of the evolution of global quantities such as the volume and mass filling fractions of different structures. For the problem of web-type classification, the results described in this work constitute the first connection between theory and observations at non-linear scales including a physical model of structure formation and the demonstrated capability of uncertainty quantification. A connection between cosmology and information theory using real data also naturally emerges from our probabilistic approach. Our results constitute quantitative chrono-cosmography of the complex web-like patterns underlying the observed galaxy distribution.
\end{abstract}


\maketitle



\section{Introduction}

The large-scale distribution of matter in the Universe is known to form intricate, complex patterns traced by galaxies. The existence of this large-scale structure (LSS), also known as the \textit{cosmic web} \citep{Bond1996}, has been suggested by early observational projects aiming at mapping the Universe \citep{Gregory1978,Kirshner1981,deLapparent1986,Geller1989,Shectman1996}, and has been extensively analyzed since then by massive surveys such as the 2dFGRS \citep{Colless2003}, the SDSS \citep[e.g.][]{Gott2005} or the 2MASS redshift survey \citep{Huchra2012}. The cosmic web is usually segmented into different elements: voids, sheets, filaments, and clusters. At late times, low-density regions (voids) occupy most of the volume of the Universe. They are surrounded by walls (or sheets) from which departs a network of denser filaments. At the intersection of filaments lie the densest clumps of matter (clusters). Dynamically, matter tends to flow out of the voids to their compensation walls, transits through filaments and finally accretes in the densest halos. 

Describing the cosmic web morphology is an involved task due to the intrinsic complexity of individual structures, but also to their connectivity and the hierarchical nature of their global organization. First approaches \citep[e.g.][]{Barrow1985,Gott1986,Babul1992,Mecke1994,Sahni1998} often characterized the LSS with a set of global and statistical diagnostics, without providing a way to locally identify cosmic web elements. In the last decade, a variety of methods has been developed for segmenting the LSS into its components and applied to numerical simulations and observations. Among them, some focus on investigating one component at a time, in particular filaments (e.g. the Candy model -- \citealp{Stoica2005,Stoica2007,Stoica2010}, the skeleton analysis -- \citealp{Novikov2006,Sousbie2008}, and DisPerSE -- \citealp{Sousbie2011a,Sousbie2011b}) or voids (e.g. \citealp{Plionis2002,Colberg2005,Shandarin2006,Platen2007,Neyrinck2008,Sutter2015VIDE,Elyiv2015}, see also \citealp{Colberg2008} for a void finder comparison project). Unfortunately, this approach does not allow an analysis of the connections between cosmic web components, identified in the same framework. Another important class of web classifiers dissects clusters, filaments, walls, and voids at the same time. In particular, several recent studies deserve special attention due to their methodological richness. The ``T-web'' and ``V-web'' \citep{Hahn2007a,Forero-Romero2009,Hoffman2012} characterize the cosmic web based on the tidal and velocity shear fields. \textsc{diva} \citep{Lavaux2010} rather uses the shear of the Lagrangian displacement field. \textsc{origami} \citep{Falck2012} identifies single and multi-stream regions in the full six-dimensional phase-space information \citep{Abel2012,Neyrinck2012,Shandarin2012}. The Multiscale Morphology Filter \citep{Aragon-Calvo2007} and later refinements \textsc{nexus}/\textsc{nexus+} \citep{Cautun2013} follow a multiscale approach which probes the hierarchical nature of the cosmic web.

In the standard theoretical picture, the cosmic web arises from the anisotropic nature of gravitational collapse, which drives the formation of structure in the Universe from primordial fluctuations \citep{Peebles1980}. The capital importance of the large-scale tidal field in the formation and evolution of the cosmic web was first pointed out in the seminal work of \citet{Zeldovich1970}. In the \citeauthor{Zeldovich1970} approximation, the late-time morphology of structures is linked to the eigenvalues of the tidal tensor in the initial conditions. Gravitational collapse amplifies any anisotropy present in the primordial density field to give rise to highly asymmetrical structures. This picture explains the segmented nature of the LSS, but not its connectivity. The cosmic web theory of \cite{Bond1996} asserted the deep connection between the tidal field around rare density peaks in the initial fluctuations and the final web pattern, in particular the filamentary cluster-cluster bridges. More generally, the shaping of the cosmic web through gravitational clustering is essentially a deterministic process described by Einstein's equations and the main source of stochasticity in the problem enters in the generation of initial conditions, which are known, from inflationary theory, to resemble a Gaussian random field to very high accuracy \citep{Guth1982,Hawking1982,Bardeen1983}. For these reasons, considerable effort has been devoted to a theoretical understanding of the LSS in terms of perturbation theory in the Eulerian and Lagrangian frames \citep[for a review, see][]{Bernardeau2002}. While this approach offers important analytical insights, it only permits to describe structure formation in the linear and mildly non-linear regimes and it is usually limited to the first few correlation functions of the density field. The complete description of the connection between primordial fluctuations and the late-time LSS, including a full phase-space treatment and the entire hierarchy of correlators, has to rely on a numerical treatment through $N$-body simulations. The characterization of cosmic web environments in the non-linear regime and the description of their time evolution has only been treated recently, following the application of web classifiers to state-of-the-art simulations. In particular, \citet{Hahn2007a,Aragon-Calvo2010} presented a local description of structure types in high-resolution cosmological simulations. \citet{Hahn2007b,Bond2010,Cautun2014} analyzed the time evolution of the cosmic web in terms of the mass and volume content of web-type components, their density distribution, and a set of new analysis tools especially designed for particular elements.

To the best of our knowledge, neither the classification of cosmic environments at non-linear scales in physical realizations of the LSS nor the investigation of their genesis and growth, using real data and with demonstrated capability of uncertainty quantification, have been treated in the existing literature. In this work, we propose the first probabilistic web-type analysis conducted with observational data in the deeply non-linear regime of LSS formation. We build accurate maps of dynamic cosmic web components with a resolution of around 3~Mpc/$h$, constrained by observations. In addition, our approach leads to the first quantitative inference of the formation history of these environments and allows the construction of maps of the embryonic traces in the initial perturbations of the late-time morphological features of the cosmic web.

Cosmographic descriptions of the LSS in terms of three-dimensional maps, and in particular  a dynamic structure type cartography carry potential for a rich variety of applications. Such maps characterize the anisotropic nature of gravitational structure formation, the clustering behavior of galaxies as a function of their tidal environment and permit to describe the traces of the cosmic web already imprinted in the initial conditions. So far, most investigations focused on understanding the physical properties of dark halos and galaxies in relation to the LSS. \citet{Hahn2007a,Hahn2007b,Hahn2009,Hahn2014,Aragon-Calvo2010} found a systematic dependence of halo properties such as morphological type, color, luminosity and spin parameter on their cosmic environment (local density, velocity and tidal field). In addition, a correlation between halo shapes and spins and the orientations of nearby filaments and sheets, predicted in simulations \citep{Altay2006,Hahn2007a,Hahn2007b,Hahn2009,Paz2008,Zhang2009,Codis2012,Libeskind2013,Welker2014,Aragon-Calvo2014,Laigle2015}, has been confirmed by observational galaxy data \citep{Paz2008,Jones2010,Tempel2013,Zhang2013}. Cartographic descriptions of the cosmic web also permit to study the environmental dependence of galaxy properties \cite[see e.g.][]{Lee2008a,Lee2008,Park2010,Yan2012,Kovavc2014} and to make the connection between the sophisticated predictions for galaxy properties in hydrodynamic simulations \citep[e.g.][]{Vogelsberger2014,Dubois2014,Codis2015} and observations. Another wide range of applications of structure type reconstructions is to probe the effect of the inhomogeneous large-scale structure on photon properties and geodesics. For example, it is possible to interpret the weak gravitational lensing effects of voids \citep{Melchior2014,Clampitt2014}. Dynamic information can also be used to produce prediction templates for secondary effects expected in the cosmic microwave background such as the kinetic Sunyaev-Zel'dovich effect \citep{Li2014}, the integrated Sachs-Wolfe and Rees-Sciama effects \citep[e.g.][]{Cai2010,Ilic2013,PlanckCollaboration2014ISW}.

Building such refined cosmographic descriptions of the Universe requires high-dimensional, non-linear inferences. In \citet{Jasche2015BORGSDSS} (\citetalias{Jasche2015BORGSDSS} in the following), we presented a chrono-cosmography project, aiming at reconstructing simultaneously the density distribution, the velocity field and the formation history of the LSS from galaxies. To do so, we used an advanced Bayesian inference algorithm to assimilate the Sloan Digital Sky Survey (SDSS) DR7 data into the forecasts of a physical model of structure formation (second order Lagrangian perturbation theory -- 2LPT). Besides inferring the four-dimensional history of the matter distribution, these results permit us an analysis of the genesis and growth of the complex web-like patterns that have been observed in our Universe. Therefore, this work constitutes a new chrono-cosmography project, aiming at the analysis of the evolving cosmic web.

Our investigations rely on the inference of the initial conditions in the SDSS volume \citepalias{Jasche2015BORGSDSS}. Starting from these, we generate a large set of constrained realizations of the Universe using the {\cola} method \citep{Tassev2013}. This physical model allows us to perform the first description of the cosmic web in the non-linear regime, using real data, and to follow the time evolution of its constituting elements. Throughout this paper, we adopt the \citet{Hahn2007a} dynamic web classifier, which segments the LSS into voids, sheets, filaments, and clusters. This choice is motivated by the close relation between the equations that dictate the dynamics of the growth of structures in the \citeauthor{Zeldovich1970} formalism and the Lagrangian description of the LSS which naturally emerges with {\borg}. As this procedure relies on the estimation of the eigenvalues of the tidal tensor in Fourier space, it constitutes a non-linear and non-local estimator of structure types, requiring adequate means to propagate observational uncertainties to finally inferred products (web-type maps and all derived quantities), in order not to misinterpret results. The {\borg} algorithm naturally addresses this problem by providing a set of density realizations constrained by the data. The variation between these samples constitute a thorough quantification of uncertainty coming from all observational effects (in particular the incompleteness of the data because of the survey mask and the radial selection functions, as well as luminosity-dependent galaxy biases, see \citetalias{Jasche2015BORGSDSS} for details), not only with a point estimation but with a detailed treatment of the likelihood. Hence, for all problems addressed in this work, we get a fully probabilistic answer in terms of a prior and a posterior distribution. Building upon the robustness of our uncertainty quantification procedure, we are able to make the first observationally-supported link between cosmology and information theory \citep[see][for theoretical considerations related to this question]{Neyrinck2014} by looking at the entropy and Kullback-Leibler divergence of probability distribution functions.

This paper is organized as follows. In section \ref{sec:Methods}, we describe our methodology: Bayesian large-scale structure inference with the {\borg} algorithm, non-linear filtering of samples with {\cola} and web-type classification using the \citeauthor{Hahn2007a} procedure. In sections \ref{sec:The late-time large-scale structure} and \ref{sec:The primordial large-scale structure}, we describe the cosmic web at present and primordial times, respectively. In section \ref{sec:Evolution of the cosmic web}, we follow the time evolution of web-types as structures form in the Universe. Finally, we summarize our results and offer concluding comments in section \ref{sec:Conclusion}.

\section{Methods}
\label{sec:Methods}

In this section, we describe our methodology step by step:

\begin{enumerate}
\item inference of the initial conditions with {\borg} (section \ref{sec:Bayesian large-scale structure inference with BORG}),
\item generation of data-constrained realizations of the SDSS volume via non-linear filtering of {\borg} samples with {\cola} (section \ref{sec:Non-linear filtering of samples with COLA}),
\item classification of the cosmic web in voids, sheets, filaments, and clusters, using the \citeauthor{Hahn2007a} algorithm (section \ref{sec:Classification of the cosmic web}).
\end{enumerate}

\subsection{Bayesian large-scale structure inference with BORG}
\label{sec:Bayesian large-scale structure inference with BORG}

This work builds upon results previously obtained by application of the {\borg} \citep[Bayesian Origin Reconstruction from Galaxies,][]{Jasche2013BORG} algorithm to Sloan Digital Sky Survey data release 7 data \citepalias{Jasche2015BORGSDSS}. {\borg} is a full-scale Bayesian inference code which permits to simultaneously analyze morphology and formation history of the cosmic web.

As discussed in \citet{Jasche2013BORG}, accurate and detailed cosmographic inferences from observations require modeling the mildly non-linear and non-linear regime of the presently observed matter distribution. The exact statistical behavior of the LSS in terms of a full probability distribution function (pdf) for non-linearly evolved density fields is not known. For this reason, the first full-scale reconstructions relied on phenomenological approximations, such as multivariate Gaussian or log-normal distributions, incorporating a cosmological power-spectrum to accurately represent correct two-point statistics of density fields \cite[see e.g.][]{Lahav1994,Zaroubi2002,Erdovgdu2004,Kitaura2008,Kitaura2009,Kitaura2010,JascheKitaura2010,Jasche2010a,Jasche2010b}. However, these prescriptions only model the one and two-point statistics of the matter distribution. Additional statistical complexity of the evolved density field arises from the fact that gravitational structure formation introduces mode coupling and phase correlations. This manifests itself not only in a sheer amplitude difference of density and velocity fields at different redshifts, but also in a modification of their statistical behavior by the generation of higher-order correlation functions. An accurate modeling of these high-order correlators is of crucial importance for a precise description of the connectivity and hierarchical nature of the cosmic web, which is the aim of this paper.

While the statistical nature of the late-time density field is poorly understood, the initial conditions from which it formed are known to obey Gaussian statistics to very great accuracy \citep{PlanckCollaboration2014PNG}. Therefore, it is reasonable to account for the increasing statistical complexity of the evolving matter distribution by a dynamical model of structure formation linking initial and final conditions. This naturally turns the problem of LSS analysis to the task of inferring the initial conditions from present cosmological observations \citep{Jasche2013BORG,Kitaura2013,Wang2013}. This approach yields a very high-dimensional and non-linear inference problem. Typically, the parameter space to explore comprises on the order of $10^6$ to $10^7$ elements, corresponding to the voxels of the map to be inferred. For reasons linked to computational cost, the {\borg} algorithm employs second order Lagrangian perturbation theory (2LPT) as an approximation for the actual gravitational dynamics linking initial three-dimensional Gaussian density fields to present, non-Gaussian density fields. As known from perturbation theory \citep[see e.g.][]{Bernardeau2002}, in the linear and mildly non-linear regime, 2LPT correctly describes the one, two and three-point statistics of the matter distribution and also approximates very well higher-order correlators. It accounts in particular for tidal effects in its regime of validity. Consequently, the {\borg} algorithm correctly transports the observational information corresponding to complex web-like features from the final density field to the corresponding initial conditions. Note that such an explicit Bayesian forward-modeling approach is always more powerful than constraining (part of) the sequence of correlation functions, as it accounts for the entire dark matter dynamics (in particular for the infinite hierarchy of correlators), in its regime of validity. This is of particular importance, since the hierarchy of correlation functions has been shown to be an insufficient description of density fields in the non-linear regime \citep{Carron2012a,Carron2012b}.

As discussed in \citetalias{Jasche2015BORGSDSS}, our analysis comprehensively accounts for observational effects such as selection functions, survey geometry, luminosity-dependent galaxy biases and noise. Corresponding uncertainty quantification is provided by sampling from the high-dimensional posterior distribution via an efficient implementation of the Hamiltonian Markov Chain Monte Carlo method \citep[see][for details]{Jasche2013BORG}. In particular, luminosity-dependent galaxy biases are explicitly part of the {\borg} likelihood and the bias amplitudes are inferred self-consistently during the run. Though not explicitly modeled, redshift-space distortions are automatically mitigated: due to the prior preference for homogeneity and isotropy, such anisotropic features are treated as noise in the data.

In the following, we make use of the 12,000 samples of the posterior distribution for primordial density fields, obtained in \citetalias{Jasche2015BORGSDSS}. These reconstructions, constrained by SDSS observations, act as initial conditions for the generation of constrained large-scale structure realizations. It is important to note that we directly make use of {\borg} outputs without any further post-processing, which demonstrates the remarkable quality of our inference results. 

\subsection{Non-linear filtering of samples with COLA}
\label{sec:Non-linear filtering of samples with COLA}

\citet[section 2.A]{Leclercq2013} performed a study of differences in the representation of structure types in density fields predicted by LPT and $N$-body simulations. To do so, they used the same web-type classification procedure as in this work (see section \ref{sec:Classification of the cosmic web}). In spite of the visual similarity of LPT and $N$-body density fields at large and intermediate scales (above a few Mpc/$h$), they found crucial differences in the representation of structures. Specifically, LPT predicts fuzzier halos than full gravity, and incorrectly assigns the surroundings of voids as part of them. This manifests itself in an overprediction of the volume occupied by clusters and voids at the detriment of sheets and filaments. The substructure of voids is also known to be incorrectly represented in 2LPT \citep{Neyrinck2013a,Leclercq2013}.

For these reasons, in this work we cannot directly make use of the final {\borg} density samples, which are a prediction of the 2LPT model. Instead, we rely on the inferred initial conditions, which contain the data constraints \citepalias[as described in][]{Jasche2015BORGSDSS} and on a non-linear filtering step similar to the one described in \citet[section 2.A]{Leclercq2015DMVOIDS}. Due to the large number of samples to be processed for this work, we do not use a fully non-linear simulation code as in \cite{Leclercq2015DMVOIDS}, but the {\cola} method \cite[COmoving Lagrangian Acceleration,][]{Tassev2013}. 

The initial density field, defined on a cubic equidistant grid with side length of 750~Mpc/$h$ and $256^3$ voxels, is populated by $512^3$ particles placed on a regular Lagrangian lattice. The particles are evolved with 2LPT to the redshift of $z=69$ and with {\cola} from $z=69$ to $z=0$. The final density field is constructed by binning the particles with a cloud-in-cell (CiC) method on a $256^3$-voxel grid. This choice corresponds to a resolution of around 3~Mpc/$h$ for all the maps described in this paper. In this fashion, we generate a large set of data-constrained reconstructions of the present-day dark matter distribution \citep[see also][]{Lavaux2010a,Kitaura2013,Hess2013,Nuza2014}. To ensure sufficient accuracy, 30 timesteps logarithmically-spaced in the scale factor are used for the evolution with \cola. We checked that this setup yields vanishing difference in the representation of final density fields with respect to the prediction of \textsc{gadget-2} \citep{Springel2001,Springel2005}. Therefore, {\cola} enables us to cheaply generate non-linear density fields at the required accuracy.

\begin{figure}
\begin{center}
\includegraphics[width=\columnwidth]{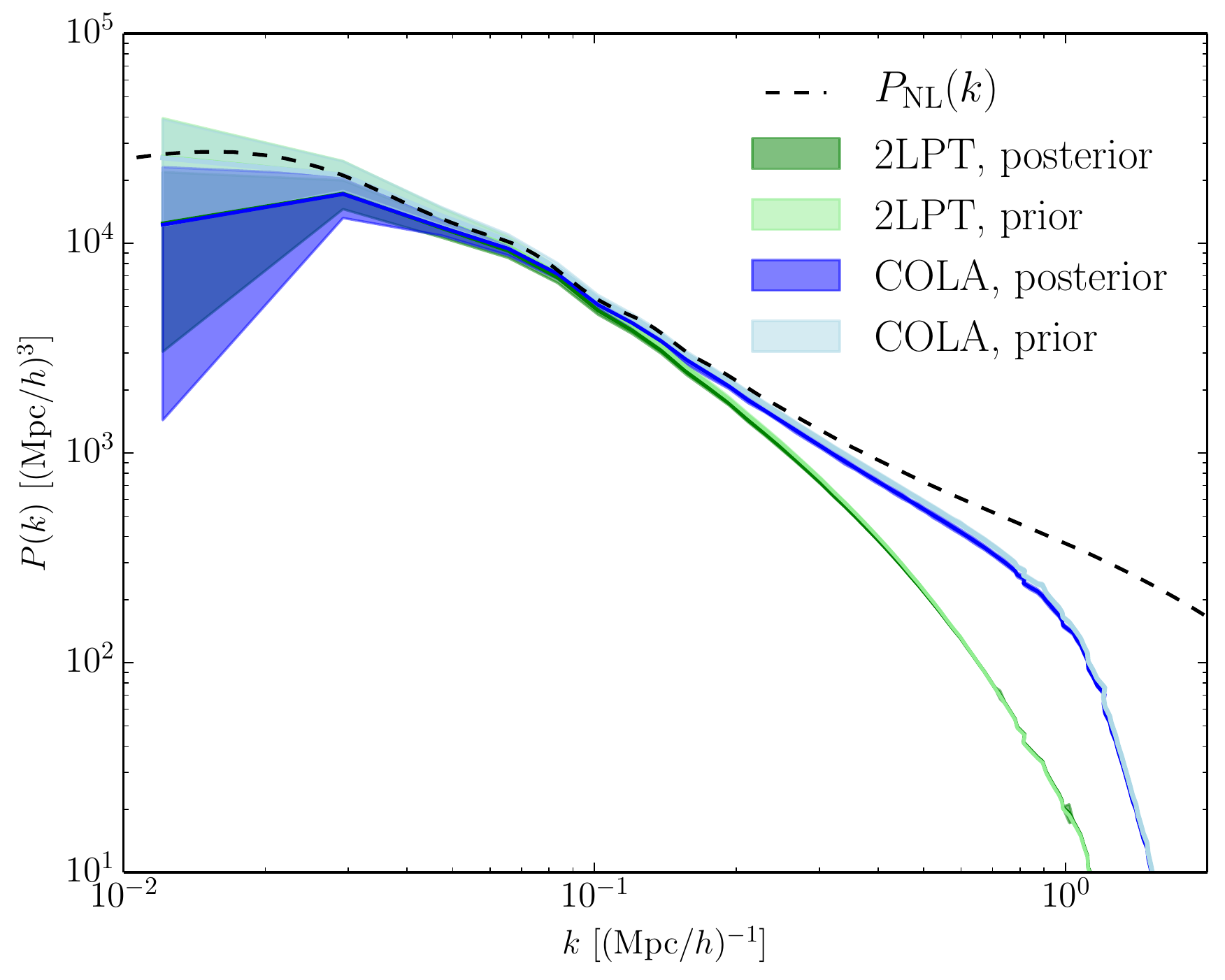}
\caption{Power spectra of dark matter density fields at redshift zero, computed with a mesh size of 3 Mpc/$h$. The particle distributions are determined using: unconstrained 2LPT realizations (``2LPT, prior''), constrained 2LPT samples inferred by {\borg} (``2LPT, posterior''), unconstrained {\cola} realizations (``COLA, prior''), constrained samples inferred by {\borg} and filtered with {\cola} (``COLA, posterior''). The solid lines correspond to the mean among all realizations used in this work, and the shaded regions correspond to the 2-$\sigma$ credible interval estimated from the standard error of the mean. The dashed black curve represents $P_\mathrm{NL}(k)$, the theoretical power spectrum expected at z = 0 from high-resolution $N$-body simulations.}
\label{fig:power_spectrum}
\end{center}
\end{figure}

As an illustration of the improvement introduced by non-linear filtering at the level of two-point statistics, in figure \ref{fig:power_spectrum}, we plot the power spectra of redshift-zero dark matter density fields. The agreement between unconstrained and constrained realizations at all scales can be checked. The plot also shows that our set of constrained reconstructions contain the additional power expected in the non-linear regime\footnote{Note that the lack of small scale power in {\cola} with respect to theoretical predictions, for $k \gtrsim 0.5~(\mathrm{Mpc}/h)^{-1}$, is a gridding artifact due to the finite mesh size used for the analysis. This value corresponds to around one quarter of the Nyquist wavenumber.}, up to $k \approx 0.4~(\mathrm{Mpc}/h)^{-1}$. For a visual illustration of the non-linear filtering procedure, which permits to check the phases of corresponding 2LPT and {\cola} fields, the reader is referred to figure 2 in \cite{Leclercq2015DMVOIDS}. The final density field predicted by {\cola} is visually indistinguishable from the right panel there, corresponding to the \textsc{gadget-2} result.

\subsection{Classification of the cosmic web}
\label{sec:Classification of the cosmic web}

The {\borg} filtered reconstructions permit a variety of scientific analyses of the large scale structure in the observed Universe. In this work, we focus specifically on the possibility to characterize the cosmic web by distinct structure types. Generally, any of the methods cited in the introduction can be employed for analysis of our density samples, however for the purpose of this paper, we follow the dynamical cosmic web classification procedure as proposed by \cite{Hahn2007a}. In analogy with the \cite{Zeldovich1970} theory, they propose to classify the large scale structure environment into four web types (voids, sheets, filaments, and clusters) based on a local-stability criterion for the orbits of test particles. The basic idea of this dynamical classification approach is that the eigenvalues of the tidal tensor characterize the geometrical properties of each point in space. The tidal tensor\footnote{We follow the nomenclature of \cite{Hahn2007a,Hoffman2012}, who call $T_{ij}$ (including its trace part) the \textit{tidal tensor}; it is called the \textit{deformation tensor} by \cite{Forero-Romero2009}.} $T_{ij}$ is given by the Hessian of the gravitational potential $\Phi$,
\begin{equation}
\label{eq:deform_tensor}
T_{ij}=\frac{\partial^2 \Phi}{\partial x_i\, \partial x_j} \, ,
\end{equation}
with $\Phi$ being the rescaled gravitational potential given by the reduced Poisson equation \cite[see appendix A in][]{Forero-Romero2009},
\begin{equation}
\label{eq:Poisson_eq}
\nabla^2 \Phi=\delta \, .
\end{equation}
With these definitions, the three eigenvalues $\lambda_1 \le \lambda_2 \le \lambda_3$ of the tidal tensor form a decomposition of the density contrast field, in the sense that the trace of $T_{ij}$ is $\lambda_1+\lambda_2+\lambda_3 = \delta$. Each spatial point can then be classified as a specific web type by considering the signs of $\lambda_1$, $\lambda_2$, $\lambda_3$. Namely, a void point corresponds to no positive eigenvalue, a sheet to one, a filament to two and a cluster to three positive eigenvalues \citep{Hahn2007a}. The interpretation of this rule is straightforward, as the sign of an eigenvalue at a given position defines whether the gravitational force in the direction of the corresponding eigenvector is contracting (positive eigenvalues) or expanding (negative eigenvalues).

Several extensions of this classification procedure exist. \cite{Forero-Romero2009} argued that rather than using a threshold value $\lambda_\mathrm{th}$ of zero, different positive values can yield better web classifications down to the megaparsec scale. \cite{Hoffman2012} reformulated the procedure using the velocity shear tensor (the ``V-web'') instead of the gravitational tidal tensor (the ``T-web''). They showed that the two classifications coincide at large scales and that the velocity field resolves finer structure than the gravitational field at the smallest scales (sub-megaparsec). In this work, we will probe scales down to $\sim$ 3~Mpc/$h$ (the voxel size in our reconstructions). Therefore, we will be content with the original classification procedure as proposed by \cite{Hahn2007a}. The structures are then classified according to the rules given in table \ref{tb:structureclassification}.

It is important to note that the tidal tensor and the rescaled gravitational potential are both physical quantities, and hence their calculation requires the availability of a full physical density field in contrast to a smoothed mean reconstruction of the density field. As described in \citetalias{Jasche2015BORGSDSS}, density samples obtained by the {\borg} algorithm provide such required full physical density fields. The tidal tensor can therefore easily be calculated in each density sample from the Fourier space representations of eq. \eqref{eq:deform_tensor} and \eqref{eq:Poisson_eq} \citep[see][for details on the technical implementation]{Hahn2007a,Forero-Romero2009}.

The aforementioned web classifiers \citep{Hahn2007a,Forero-Romero2009,Hoffman2012} provide four voxel-wise scalar fields that characterize the large scale structure. In a specific realization, the answer is unique, meaning that these fields obey the following conditions at each voxel position $\vec{x}_k$:
\begin{equation}
\mathrm{T}_i(\vec{x}_k) \in \{0,1\} \; \mathrm{for} \; i \in \llbracket 0,3 \rrbracket \quad \mathrm{and} \quad \sum_{i=0}^{3} \mathrm{T}_i(\vec{x}_k) = 1 
\end{equation}
where $\mathrm{T}_0=$ void, $\mathrm{T}_1=$ sheet, $\mathrm{T}_2=$ filament, $\mathrm{T}_3=$ cluster. In this work, we follow the Bayesian approach of \citetalias{Jasche2015BORGSDSS} and quantify the degree of belief in structure type classification. Specifically, our web classification is given in terms of four voxel-wise scalar fields that obey the following conditions at each voxel position $\vec{x}_k$:
\begin{equation}
\mathcal{T}_i(\vec{x}_k) \in [0,1] \; \mathrm{for} \; i \in \llbracket 0,3 \rrbracket \quad \mathrm{and} \quad \sum_{i=0}^{3} \mathcal{T}_i(\vec{x}_k) = 1 .
\end{equation}
Here, $\mathcal{T}_i(\vec{x}_k) \equiv \langle \mathrm{T}_i(\vec{x}_k) \rangle_{\mathcal{P}(\mathrm{T}_i(\vec{x}_k)|d)} = \mathcal{P}(\mathrm{T}_i(\vec{x}_k)|d)$ are the posterior probabilities indicating the possibility to encounter specific structure types at a given position in the observed volume, conditional on the data. These are estimated by applying the web classification to all density samples and counting the relative frequencies at each individual spatial coordinate within the set of samples \citep[see section 5 in][]{Jasche2010a}. With this definition, the cosmic web-type posterior mean is given by
\begin{equation}
\left\langle \mathcal{P}(\mathrm{T}_i(\vec{x}_k)|d) \right\rangle = \frac{1}{N}\sum_{n=1}^{N} \sum_{j=0}^{3} \delta^{(\mathrm{K})}_{\mathrm{T}_i(\vec{x}_k)\mathrm{T}^n_j(\vec{x}_k)} ,
\label{eq:pdf_mean}
\end{equation}
where $n$ labels one of the $N$ samples, $\mathrm{T}^n_j(\vec{x}_k)$ is the result of the web classifier on the $n$-th sample (i.e. a unit four-vector at each voxel position $\vec{x}_k$ containing zeros except for one component, which indicates the structure type), and $\delta^{(\mathrm{K})}$ is a Kronecker symbol.

\begin{table}\centering
\begin{tabular}{ll}
\hline\hline
Structure type & Rule\\
\hline
Void & $\lambda_1,\lambda_2,\lambda_3 < 0$\\
Sheet & $\lambda_1,\lambda_2 < 0$ and $\lambda_3 > 0$\\
Filament & $\lambda_1 < 0$ and $\lambda_2,\lambda_3 > 0$\\
Cluster & $\lambda_1,\lambda_2,\lambda_3 > 0$\\
\hline\hline
\end{tabular}
\caption{Rules for the dynamic classification of web types.}
\label{tb:structureclassification}
\end{table}

\section{The late-time large-scale structure}
\label{sec:The late-time large-scale structure}

In this section, we discuss the results of our analysis of the final density field, at $a=1$. For reasons of computational time with {\cola} filtering (see section \ref{sec:Non-linear filtering of samples with COLA}), we kept around 10\% of the original set of samples obtained in \citetalias{Jasche2015BORGSDSS}. In order to mitigate as much as possible the effects of correlation among samples, we maximally separated the samples kept for the present analysis, keeping one out of ten consecutive samples of the original Markov Chain. Hence,  for all results discussed in this section, we used a total of 1,097 samples inferred by {\borg} and filtered with {\cola}.

\subsection{Tidal environment}
\label{sec:Tidal environment final}

\begin{figure*}
\begin{center}
\includegraphics[width=\textwidth]{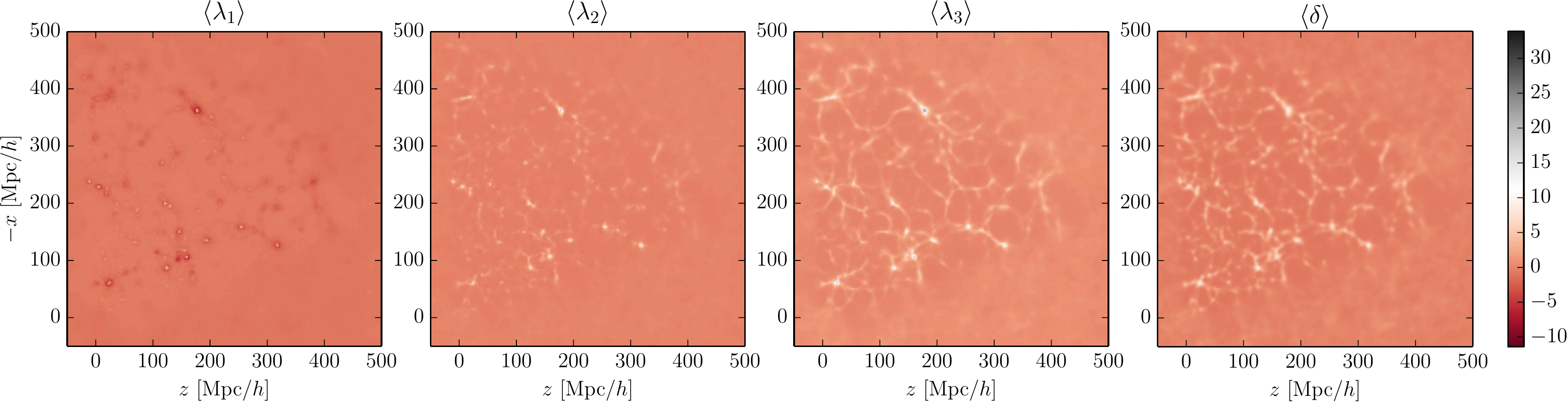}
\caption{Slices through the three-dimensional ensemble posterior mean for the eigenvalues $\lambda_1 \leq \lambda_2 \leq \lambda_3$ of the tidal field tensor in the final conditions, estimated from 1,097 samples. The rightmost panel shows the corresponding slice through the posterior mean for the final density contrast $\delta=\lambda_1+\lambda_2+\lambda_3$, obtained in \cite{Jasche2015BORGSDSS}. \label{fig:lambda_cola_final_mean}}
\end{center}
\end{figure*}

\begin{figure*}
\begin{center}
\includegraphics[width=\textwidth]{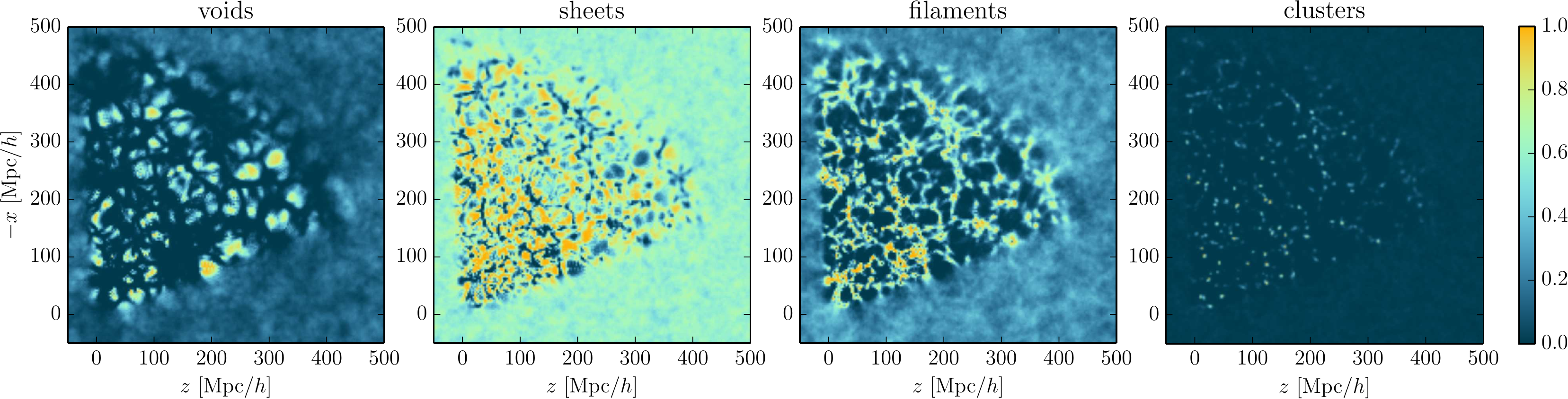}
\caption{Slices through the posterior mean for different structure types (from left to right: void, sheet, filament, and cluster) in the late-time large-scale structure in the Sloan volume ($a=1$). These four three-dimensional voxel-wise pdfs sum up to one on a voxel basis.\label{fig:pdf_final}}
\end{center}
\end{figure*}

As a natural byproduct, the application of the \citeauthor{Hahn2007a} web classifier to density samples yields samples of the probability distribution functions for the three eigenvalues of the tidal field tensor. These pdfs account for the assumed physical model of structure formation and the data constraints, and quantify uncertainty coming in particular from selection effects, surveys geometries and galaxy biases. In a similar fashion as described in \citetalias{Jasche2015BORGSDSS}, the ensemble of samples permits us to provide any desired statistical summary such as mean and variance.

In figure \ref{fig:lambda_cola_final_mean}, we show slices\footnote{In all slice plots of this paper, we kept the coordinate system of \citetalias{Jasche2015BORGSDSS}.} through the ensemble mean fields $\lambda_1$, $\lambda_2$ and $\lambda_3$. For visual comparison, the rightmost panel of figure \ref{fig:lambda_cola_final_mean} shows the corresponding slice through the posterior mean of the final density contrast, $\delta = \lambda_1 + \lambda_2 + \lambda_3$, obtained in \citetalias{Jasche2015BORGSDSS}. Different morphologies can be observed in the data-constrained parts of these slices: $\lambda_1$, $\lambda_2$ and $\lambda_3$ respectively trace well the clusters, filaments and sheets, as we now argue. The $\lambda_1$ field is rather homogeneous, apart for small spots where all eigenvalues are largely positive, i.e. undergoing dramatic gravitational collapse along three axes. These correspond to the dynamic clusters. Note that there exists a form of ``tidal compensation'': these clusters are surrounded by regions where $\lambda_1$ is smaller than its cosmic mean. More patterns can be observed in the $\lambda_2$ field: it also exhibits filaments (appearing as dots when piercing the slice). Finally, the $\lambda_3$ field is highly-structured, as it also traces sheets (which appear filamentary when sliced). Dynamic voids can also be easily distinguished in this field, wherever $\lambda_3$ is negative.

\subsection{Probabilistic web-type cartography}
\label{sec:Probabilistic web-type cartography final}

Building upon previous results and using the procedure described in section \ref{sec:Classification of the cosmic web}, we obtain probabilistic maps of structures. More precisely, we obtain four probability distributions at each spatial position, $\mathcal{P}(\mathrm{T}_i(\vec{x_k})|d)$, indicating the possibility to encounter a specific structure type (cluster, filament, sheet, void) at that position. As noted in section \ref{sec:Classification of the cosmic web}, these pdfs take their values in the range $[0,1]$ and sum up to one on a voxel-basis. Figure \ref{fig:pdf_final} shows slices through their means (see equation \eqref{eq:pdf_mean}). The plot shows the anticipated behavior, with a high degree of structure and values close to certainty (i.e. zero or one) in regions covered by data, while the unobserved regions approach a uniform value corresponding to the prior. At this point, it is worth noting that the \citeauthor{Hahn2007a} web classifier has a prior preference for some structure types. Using unconstrained large-scale structure realizations produced with the same setup\footnote{By this, we specifically mean realizations obtained from initial randomly-generated Gaussian density fields with an \cite{Eisenstein1998,Eisenstein1999} power spectrum using the fiducial cosmological parameters of the {\borg} analysis ($\Omega_\mathrm{m}~=~0.272$, $\Omega_\mathrm{\Lambda}~=~0.728$, $\Omega_\mathrm{b}~=~0.045$, $h~=~0.702$, $\sigma_8~=~0.807$, $n_\mathrm{s}~=~0.961$, see \citetalias{Jasche2015BORGSDSS}\label{footnote:setup}). The density field is defined on a $750~\mathrm{Mpc}/h$ cubic grid of $256^3$-voxels and populated by $512^3$ dark matter particles, which are evolved to $z=69$ with 2LPT and from $z=69$ to $z=0$ with {\cola}, using 30 timesteps logarithmically-spaced in the scale factor. The particles are binned on a $256^3$-voxels grid with the CiC scheme to get the final density field.}, we measured that these prior probabilities, $\mathcal{P}(\mathrm{T}_i)$, can be well described by Gaussians whose mean and standard deviation are given in table \ref{tb:prior_final}. 

\begin{table}\centering
\begin{tabular}{lcc}
\hline\hline
Structure type & $\mu_{\mathcal{P}(\mathrm{T}_i)}$ & $\sigma_{\mathcal{P}(\mathrm{T}_i)}$ \\
\hline
\multicolumn{3}{c}{Late-time large-scale structure ($a=1$)} \\
Void & $0.14261$ & $6.1681 \times 10^{-4}$ \\
Sheet & $0.59561$ & $6.3275 \times 10^{-4}$ \\
Filament & $0.24980$ & $5.5637 \times 10^{-4}$ \\
Cluster & $0.01198$ & $5.8793 \times 10^{-5}$ \\
\hline\hline
\end{tabular}
\caption{Prior probabilities assigned by the \citet{Hahn2007a} web classifier to the different structures types, in the late-time large-scale structure ($a=1$).}
\label{tb:prior_final}
\end{table}

In addition to their ensemble mean, the set of samples permits to propagate uncertainty quantification to web-type classification. In particular, it allows us to locally assess the strength of data constraints. In information theory, a convenient way to characterize the uncertainty content of a random source $\mathcal{S}$ is the Shannon entropy \citep{Shannon1948}, defined by
\begin{equation}
H\left[\mathcal{S}\right] \equiv - \sum_i p_i \log_2(p_i) ,
\end{equation}
where the $p_i$ are the probabilities of possible events. This definition yields expected properties and accounts for the intuition that the more likely an event is, the less information it provides when it occurs (i.e. the more it contributes to the source entropy). We follow this prescription and write the voxel-wise entropy of the web-type posterior, $\mathcal{P}(\mathrm{T}(\vec{x}_k)|d)$, as
\begin{equation}
H\left[ \mathcal{P}(\mathrm{T}(\vec{x}_k)|d) \right] \equiv - \sum_{i=0}^{3} \mathcal{P}(\mathrm{T}_i(\vec{x}_k)|d) \log_2(\mathcal{P}(\mathrm{T}_i(\vec{x}_k)|d)) .
\label{eq:definition_entropy}
\end{equation}
It is a number in the range $[0,2]$ and its natural unit is the shannon (Sh). $H=0$~Sh in the case of perfect certainty, i.e. when the data constraints entirely determine the underlying structure type: $\mathcal{P}(\mathrm{T}_{i_0}(\vec{x}_k)|d)$ is 1 for one $i_0$ and 0 for $i \neq i_0$. $H$ reaches its maximum value of $2$~Sh when all $\mathcal{P}(\mathrm{T}_i(\vec{x}_k)|d)$ are equal to $1/4$. This is the case of maximal randomness: all the events being equally likely, no information is gained when one occurs.

\begin{figure*}
\begin{center}
\includegraphics[width=0.49\textwidth]{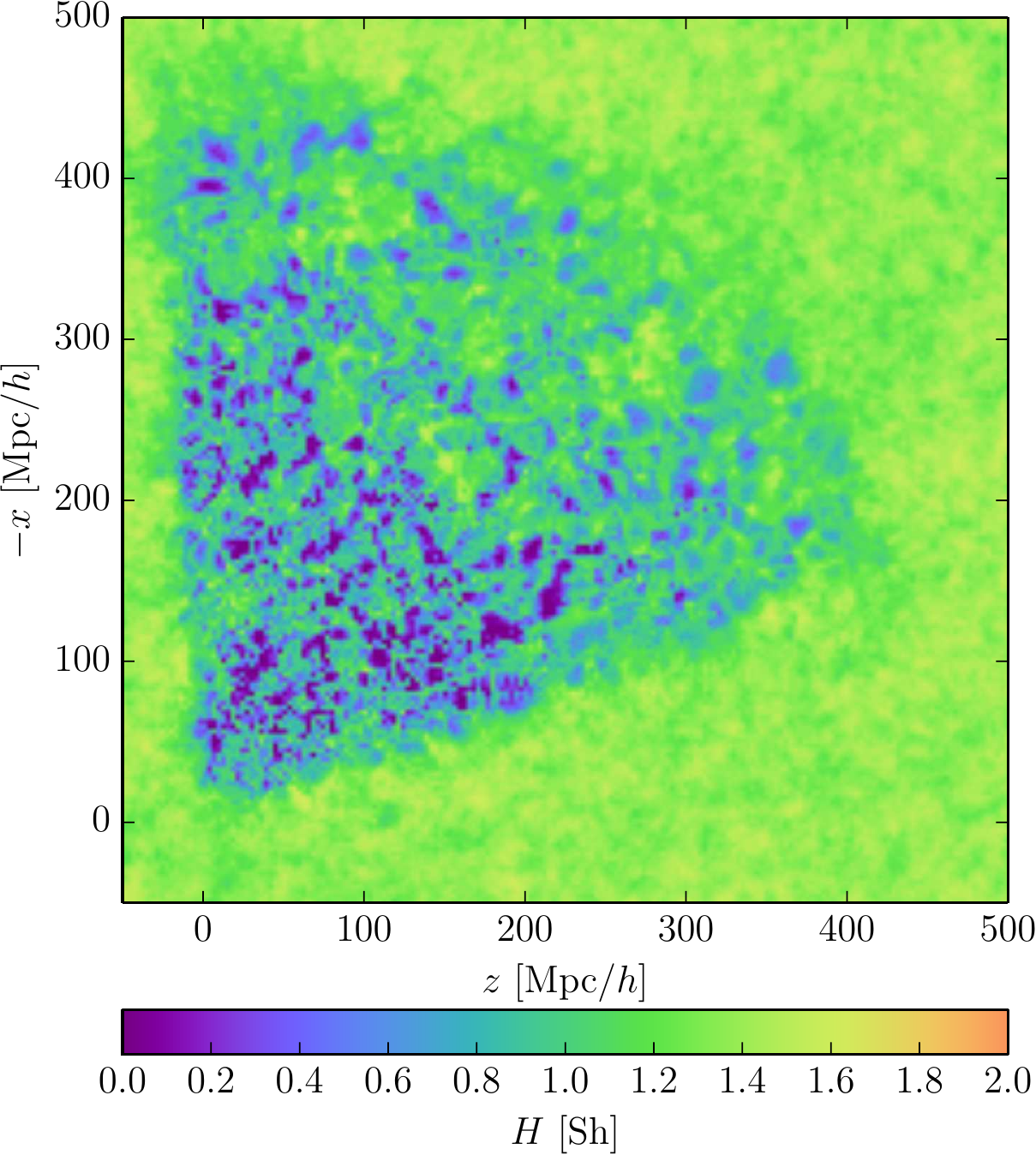}
\includegraphics[width=0.49\textwidth]{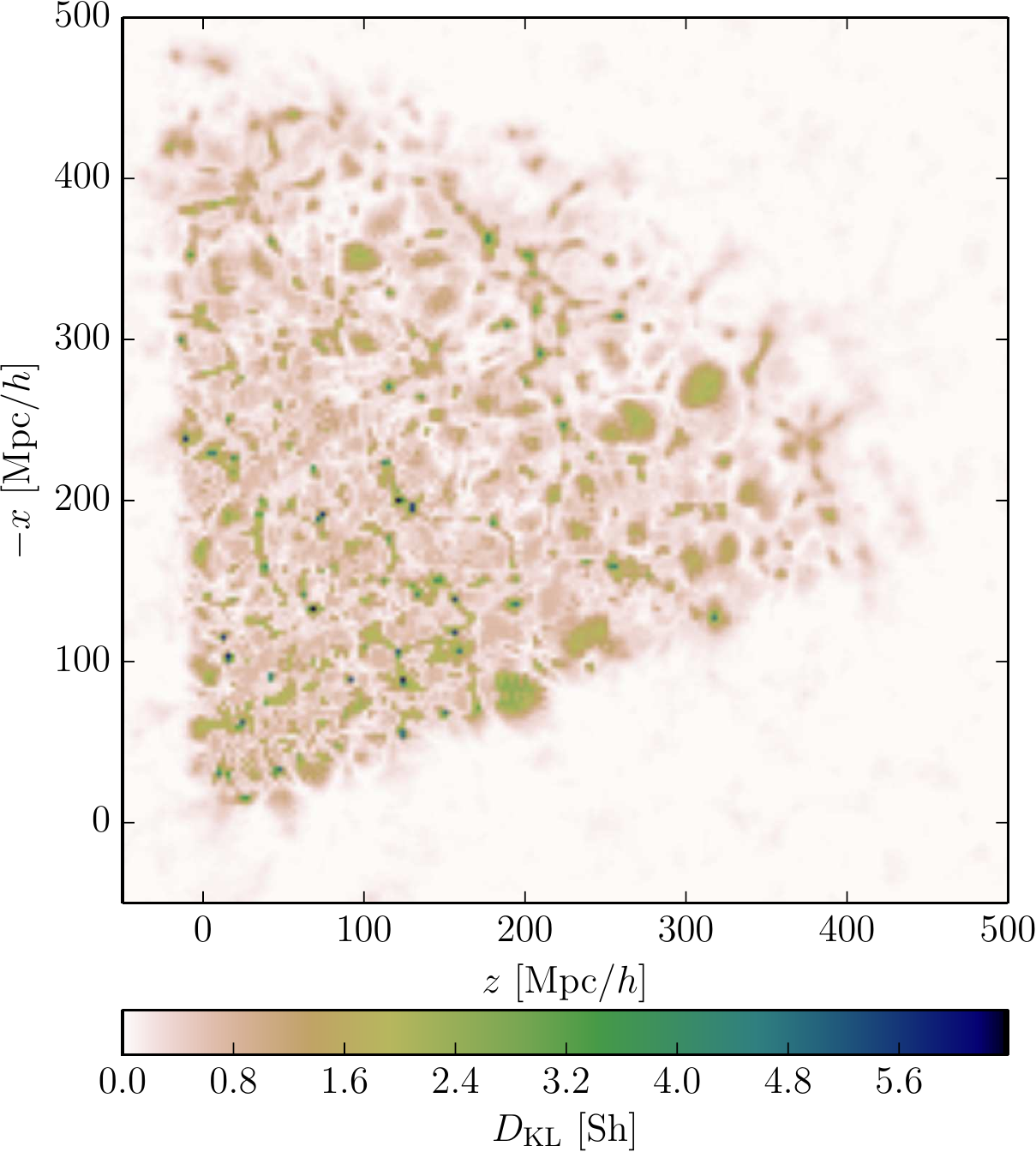}
\caption{Slices through the entropy of the structure types posterior (left panel) and the Kullback-Leibler divergence of the posterior from the prior (right panel), in the final conditions. The entropy $H$, defined by equation \eqref{eq:definition_entropy}, quantifies the information content of the posterior pdf represented in figure \ref{fig:pdf_final}, which results from fusing the information content of the prior and the data constraints. The Kullback-Leibler divergence $D_\mathrm{KL}$, defined by equation \eqref{eq:KL_divergence}, represents the information gained in moving from the prior to the posterior. It quantifies the information that has been learned on structure types by looking at SDSS galaxies.\label{fig:pdf_final_entropy}}
\end{center}
\end{figure*}

A slice through the voxel-wise entropy of the web-type posterior is shown in the left panel of figure \ref{fig:pdf_final_entropy}. Generally, the entropy map reflects the information content of the posterior pdf, which comes from augmenting the information content of the prior pdf with the data constraints, in the Bayesian way.

The entropy takes low values and shows a high degree of structure in the regions where data constraints exist, and even reaches zero in some spots where the data are perfectly informative. Comparing with figures \ref{fig:lambda_cola_final_mean} and \ref{fig:pdf_final}, one can note that this structure is highly non-trivial and does not follow any of the previously described patterns. This is due to the facts that in a Poisson process, the signal (here the density, inferred in \citetalias{Jasche2015BORGSDSS}) is correlated with the uncertainty and that structure types classification further is a non-linear function of the density field. In the unobserved regions, the entropy fluctuates around a constant value of about $1.4$~Sh, which characterizes the information content of the prior. This value is consistent with the expectation, which can be computed using equation \eqref{eq:definition_entropy} (unconditional on the data) and the numbers given in table \ref{tb:prior_final}. 

The information-theoretic quantity that measures the information gain (in shannons) due to the data is the relative entropy or Kullback-Leibler divergence \citep{Kullback1951} of the posterior from the prior,
\begin{widetext}
\begin{equation}
D_\mathrm{KL}\left[ \mathcal{P}(\mathrm{T}(\vec{x}_k)|d) \middle\| \mathcal{P}(\mathrm{T}) \right]
\equiv \sum_{i=0}^{3} \mathcal{P}(\mathrm{T}_i(\vec{x}_k)|d) \log_2\left(\frac{\mathcal{P}(\mathrm{T}_i(\vec{x}_k)|d)}{\mathcal{P}(\mathrm{T}_i)}\right)
= - H\left[ \mathcal{P}(\mathrm{T}(\vec{x}_k)|d) \right] - \sum_{i=0}^{3} \mathcal{P}(\mathrm{T}_i(\vec{x}_k)|d) \log_2(\mathcal{P}(\mathrm{T}_i)) .
\label{eq:KL_divergence}
\end{equation}
\end{widetext}
It is a non-symmetric measure of the difference between the two probability distributions.

A slice through the voxel-wise Kullback-Leibler divergence of the web-type posterior from the prior is shown in the right panel of figure \ref{fig:pdf_final_entropy}. As expected, the information gain is zero out of the survey boundaries. In the observed regions, SDSS galaxies are informative on underlying structure types at the level of at least $\sim$~1~Sh. This number can go to $\sim$~3~Sh in the interior of deep voids and up to $\sim$~6~Sh in the densest clusters. This map permits to visualize the regions where additional data would be needed to improve structure type classification, e.g. in some high-redshift regions where uncertainty remains due to selection effects.

\subsection{Volume and mass filling fractions}
\label{sec:Volume and mass filling fractions final}

A characterization of large scale environments commonly found in literature involves evaluating global quantities such as the volume and mass content of these structures. In a particular realization, the volume filling fraction (VFF) for structure type $\mathrm{T}_i$ is the number of voxels of type $\mathrm{T}_i$ divided by the total number of voxels in the considered volume,
\begin{equation}
\mathrm{VFF}(\mathrm{T}_i) \equiv \frac{\sum_{\vec{x}_k} \sum_{j=0}^{3} \delta^{(\mathrm{K})}_{\mathrm{T}_i(\vec{x}_k)\mathrm{T}^n_j(\vec{x}_k)}}{N_{\mathrm{vox}}} .
\label{eq:definition_VFF}
\end{equation}
The mass filling fraction (MFF) can be obtained in a similar manner by weighting the same sum by the local density $\rho(\vec{x}_k) = \bar{\rho}\,(1+\delta(\vec{x}_k))$,
\begin{equation}
\mathrm{MFF}(\mathrm{T}_i) \equiv \frac{\sum_{\vec{x}_k} \sum_{j=0}^{3} (1+\delta(\vec{x}_k)) \delta^{(\mathrm{K})}_{\mathrm{T}_i(\vec{x}_k)\mathrm{T}^n_j(\vec{x}_k)}}{\sum_{\vec{x}_k} (1+\delta(\vec{x}_k))} .
\label{eq:definition_MFF}
\end{equation}
To ensure that results are not prior-dominated, we measured the VFFs and MFFs in the data-constrained parts of our realizations. More precisely, we limited ourselves to the voxels where the survey response operator (representing simultaneously the survey geometry and the selection effects, see \citetalias{Jasche2015BORGSDSS}) is strictly positive. This amounts to $N_\mathrm{vox} = 3$,$148$,$504$ out of $256^3~=~16$,$777$,$216$ voxels, around $18.7$\% of the full box \citep[see also section II.C.2. and figure 3 in][]{Leclercq2015DMVOIDS}. In equations \eqref{eq:definition_VFF} and \eqref{eq:definition_MFF}, $\vec{x}_k$ labels one of these voxels. 

By measuring the VFF and MFF of different structure types in each constrained realization of our ensemble, we obtained the posterior pdfs, $\mathcal{P}(\mathrm{VFF}(\mathrm{T}_i)|d)$ and $\mathcal{P}(\mathrm{MFF}(\mathrm{T}_i)|d)$, conditional on the data. Similarly, we computed the prior pdfs, $\mathcal{P}(\mathrm{VFF}(\mathrm{T}_i))$ and $\mathcal{P}(\mathrm{MFF}(\mathrm{T}_i))$, using unconstrained realizations produced with the same setup. We found that all these pdfs can be well described by Gaussians, the mean and variance of which are given in tables \ref{tb:final_vff} and \ref{tb:final_mff}. 

Previous studies on this topic \citep[e.g.][]{Doroshkevich1970b,Shen2006,Hahn2007a,Forero-Romero2009,Jasche2010a,Aragon-Calvo2010,Shandarin2012,Cautun2014} have found a wide range of values for the VFF and MFF of structures \citep[see e.g. table 3 in][]{Cautun2014}. For example, existing studies found that clusters occupy at most a few percent of the volume of the Universe but contribute significantly to the mass content, with a MFF ranging from $\sim 10\%$ \citep{Hahn2007a,Cautun2014} to $\sim 40\%$ \citep{Shandarin2012}. The void volume fraction can vary from $\sim 10\%$ \citep{Hahn2007a} to $\sim 80\%$ \citep{Aragon-Calvo2010,Shandarin2012,Cautun2014}; in the \cite{Forero-Romero2009} formalism, it is a very sensitive function of the threshold $\lambda_\mathrm{th}$ \citep[figure 9 in][]{Jasche2010a}. These large disparities in the literature arise because different algorithms use various information and criteria for classifying the cosmic web. For this reason, we believe that it is only relevant to make relative statements for the same setup, i.e. to compare our results to the corresponding prior quantities, as done in tables \ref{tb:final_vff} and \ref{tb:final_mff}. In this purpose, the large number of samples used allowed a precise characterization of the pdfs so that all digits quoted in the tables are significant. Note that all our analyses are repeatable for different setups, which allows in principle a comparison with any previous work.

As expected for a Bayesian update of the degree of belief, the posterior quantities generally have smaller variance and a mean value displaced from the prior mean. For the MFF, the posterior means are always within two standard deviations of the corresponding prior means. The analysis shows that in the SDSS, a larger mass fraction is occupied by clusters, sheets, and voids, at the detriment of filaments, in comparison to the prior expectation. The data also favor a smaller filling of the Sloan volume by filaments and sheets and larger filling by voids and clusters. For the cluster VFF, the posterior mean, $\mu_{\mathrm{VFF}(\mathrm{T}_3)|d} = 0.01499$ is at about $15$ standard deviations ($\sigma_{\mathrm{VFF}(\mathrm{T}_3)}= 1.9194 \times 10^{-4}$) of the prior mean, $\mu_{\mathrm{VFF}(\mathrm{T}_3)} = 0.01198$. Given other results on the VFF and MFF, we believe that the data truly favor a higher volume content in clusters as compared to the structure formation model used as prior. However, this surprising result should be treated with care; part of the discrepancy is likely due to the original \textsc{borg} analysis, which optimizes the initial conditions for evolution with 2LPT (instead of the non-linear evolution with \textsc{cola} used for the present work). LPT predicts fuzzier halos than $N$-body dynamics, which results in the incorrect prediction of a high cluster VFF \citep{Leclercq2013}. 

\begin{table}\centering
\begin{tabular}{lcccc}
\hline\hline
Structure type & $\mu_{\mathrm{VFF}}$ & $\sigma_\mathrm{VFF}$ & $\mu_{\mathrm{VFF}}$ & $\sigma_\mathrm{VFF}$ \\
\hline
\multicolumn{1}{c}{} & \multicolumn{4}{c}{Late-time large-scale structure ($a=1$)} \\
\multicolumn{1}{c}{} & \multicolumn{2}{c}{Posterior} & \multicolumn{2}{c}{Prior} \\
Void & $0.14897$ & $1.8256 \times 10^{-3}$ & $0.14254$ & $6.2930 \times 10^{-3}$ \\
Sheet & $0.58914$ & $1.3021 \times 10^{-3}$ & $0.59562$ & $2.2375 \times 10^{-3}$ \\
Filament & $0.24689$ & $1.1295 \times 10^{-3}$ & $0.24986$ & $4.4440 \times 10^{-3}$ \\
Cluster & $0.01499$ & $8.7274 \times 10^{-5}$ & $0.01198$ & $1.9194 \times 10^{-4}$ \\
\hline\hline
\end{tabular}
\caption{Mean and standard deviation of the prior and posterior pdfs for the volume filling fraction of different structure types in the late-time large-scale structure ($a=1$).}
\label{tb:final_vff}
\end{table}

\begin{table}\centering
\begin{tabular}{lcccc}
\hline\hline
Structure type & $\mu_{\mathrm{MFF}}$ & $\sigma_\mathrm{MFF}$ & $\mu_{\mathrm{MFF}}$ & $\sigma_\mathrm{MFF}$ \\
\hline
\multicolumn{1}{c}{} & \multicolumn{4}{c}{Late-time large-scale structure ($a=1$)} \\
\multicolumn{1}{c}{} & \multicolumn{2}{c}{Posterior} & \multicolumn{2}{c}{Prior} \\
Void & $0.04050$ & $8.3531 \times 10^{-4}$ & $0.03876$ & $2.3352 \times 10^{-3}$ \\
Sheet & $0.35605$ & $1.2723 \times 10^{-3}$ & $0.35286$ & $3.6854 \times 10^{-3}$ \\
Filament & $0.47356$ & $1.5661 \times 10^{-3}$ & $0.48170$ & $4.2215 \times 10^{-3}$ \\
Halo & $0.12990$ & $6.4966 \times 10^{-4}$ & $0.12666$ & $1.8284 \times 10^{-3}$ \\
\hline\hline
\end{tabular}
\caption{Mean and standard deviation of the prior and posterior pdfs for the mass filling fraction of different structure types in the late-time large-scale structure ($a=1$).}
\label{tb:final_mff}
\end{table}

\section{The primordial large-scale structure}
\label{sec:The primordial large-scale structure}

In this section, we discuss the results of our analysis of the initial density field, at $a=10^{-3}$. Since the analysis of the primordial large-scale structure does not involve an additional filtering step, we have been able to keep a larger number of samples of the posterior pdf for initial conditions, obtained in \citetalias{Jasche2015BORGSDSS}. Hence, for all results described in this section, we used a total of 4,473 samples.

\subsection{Tidal environment}

\begin{figure*}
\begin{center}
\includegraphics[width=\textwidth]{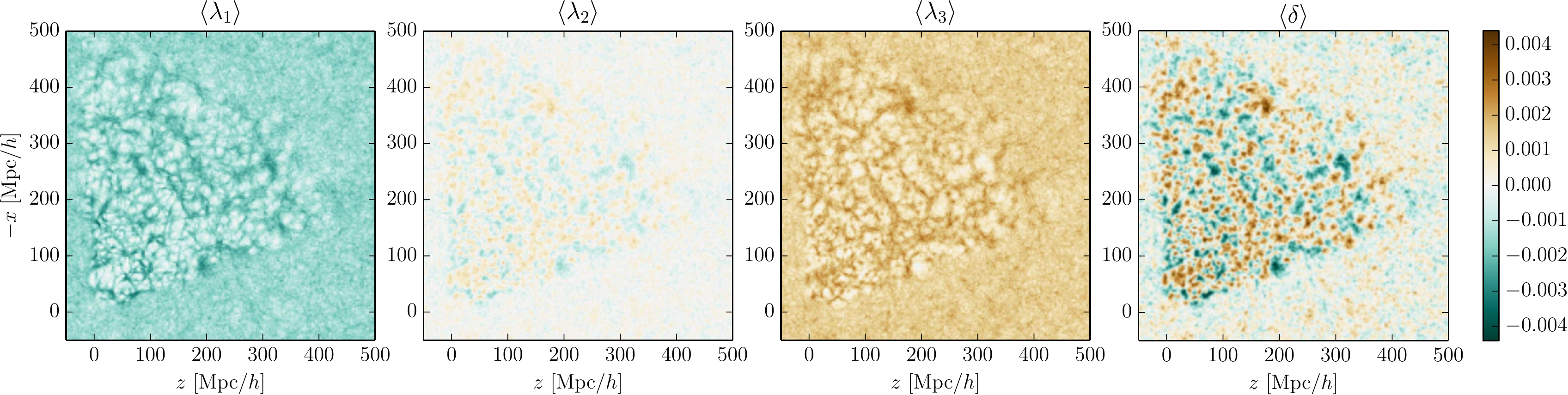}
\caption{Slices through the three-dimensional ensemble posterior mean for the eigenvalues $\lambda_1 \leq \lambda_2 \leq \lambda_3$ of the tidal field tensor in the initial conditions, estimated from 4,473 samples. The rightmost panel shows the corresponding slice through the posterior mean for the initial density contrast $\delta=\lambda_1+\lambda_2+\lambda_3$, obtained in \cite{Jasche2015BORGSDSS}. \label{fig:lambda_initial_mean}}
\end{center}
\end{figure*}

\begin{figure*}
\begin{center}
\includegraphics[width=\textwidth]{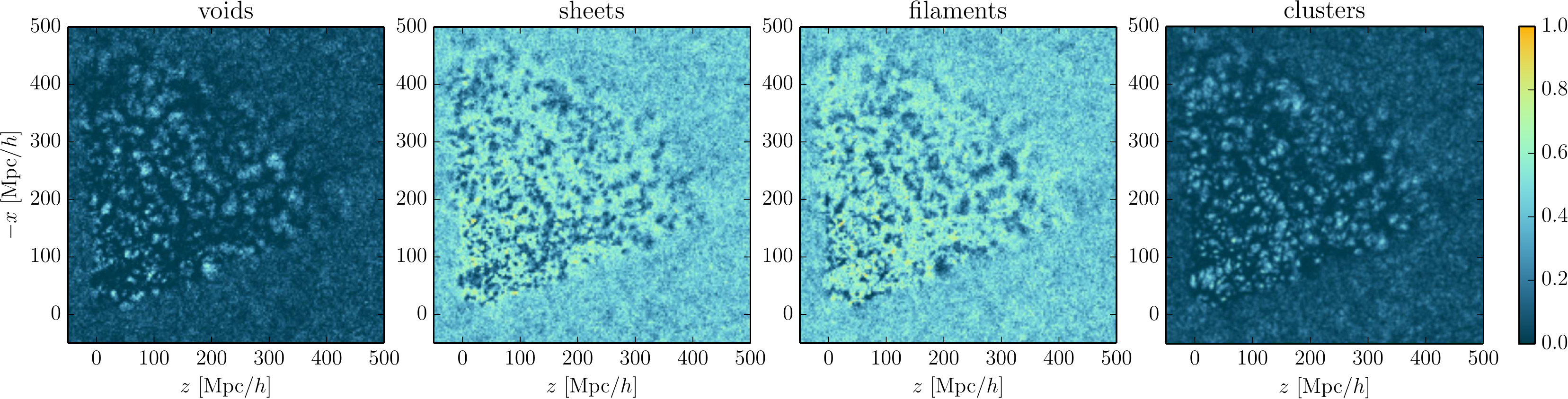}
\caption{Slices through the posterior mean for different structure types (from left to right: void, sheet, filament, and cluster) in the primordial large-scale structure in the Sloan volume ($a=10^{-3}$). These four three-dimensional voxel-wise pdfs sum up to one on a voxel basis.\label{fig:pdf_initial}}
\end{center}
\end{figure*}

In a similar fashion as in section \ref{sec:Tidal environment final}, the application of the \citeauthor{Hahn2007a} web classifier to initial density samples yields the posterior pdf for the three eigenvalues, $\lambda_1$, $\lambda_2$ and $\lambda_3$, of the initial tidal field tensor. Figure \ref{fig:lambda_initial_mean} shows slices through their means. For visual comparison, the rightmost panel shows the corresponding slice through the posterior mean of the initial density contrast, $\delta~=~\lambda_1~+~\lambda_2~+~\lambda_3$, obtained in \citetalias{Jasche2015BORGSDSS}.

In a Gaussian random field, $\lambda_1$ is generally negative, $\lambda_3$ is generally positive and $\lambda_2$ close to zero (see the unobserved parts of the slices in figure \ref{fig:lambda_initial_mean}). In addition, $\lambda_2$ closely resembles the total density contrast $\delta$ up to a global scaling. In the constrained regions, the eigenvalues of the initial tidal tensor follow this behavior. The structure observed in their maps is visually consistent with the decomposition of Gaussian density fluctuations as shown by the right panel.

\subsection{Probabilistic web-type cartography}
\label{sec:Probabilistic web-type cartography initial}

Looking at the sign of the eigenvalues of the initial tidal tensor and following the procedure described in section \ref{sec:Classification of the cosmic web}, we obtain a probabilistic cartography of the primordial large-scale structure. As before, we obtain four voxel-wise pdfs $\mathcal{P}(\mathrm{T}_i(\vec{x_k})|d)$, taking their values in the range $[0,1]$ and summing up to one. Figure \ref{fig:pdf_initial} shows slices through their means, defined by equation \eqref{eq:pdf_mean}. As in the final conditions, the maps exhibit structure in the data-constrained regions and approach uniform values in the unobserved parts, corresponding to the respective priors. Using unconstrained realizations of Gaussian random fields produced with the same setup\footnote{We used the initial conditions of our set of unconstrained simulations (see footnote \ref{footnote:setup}).}, we measured these prior probabilities. Their means and standard deviations are given in table \ref{tb:prior_initial}.

At this point, it is worth mentioning that there exists an additional symmetry for Gaussian random fields. Since the definition of the tidal tensor is linear in the density contrast (see equations \eqref{eq:deform_tensor} and \eqref{eq:Poisson_eq}) and since positive and negative density contrasts are equally likely, a positive and negative value for a given $\lambda_i$ have the same probabilities. Because of this sign symmetry, the pdfs for voids and clusters (0 or 3 positive/negative eigenvalues) and the pdfs for sheets and filaments (1 or 2 positive/negative eigenvalues) are equal. This can be checked both in table \ref{tb:prior_initial} and in the unconstrained regions of the maps in figure \ref{fig:pdf_initial}. In the constrained regions, a qualitative complementarity between pdfs for voids and clusters and for sheets and filaments can be observed. This can be explained by the following. As $\sum_i \mathcal{P}(\mathrm{T}_i(\vec{x_k})|d)=1$ and assuming that $\mathcal{P}(\mathrm{T}_i(\vec{x_k})|d) \approx \mathcal{P}(\mathrm{T}_{3-i}(\vec{x_k})|d)$ for unlikely events, consistently with the previous remark, we get $\mathcal{P}(\mathrm{T}_0(\vec{x_k})|d) \approx 1 - \mathcal{P}(\mathrm{T}_3(\vec{x_k})|d)$ wherever $\mathcal{P}(\mathrm{T}_1(\vec{x_k})|d) \approx \mathcal{P}(\mathrm{T}_2(\vec{x_k})|d)$ is sufficiently small; and $\mathcal{P}(\mathrm{T}_1(\vec{x_k})|d) \approx 1 - \mathcal{P}(\mathrm{T}_2(\vec{x_k})|d)$ wherever $\mathcal{P}(\mathrm{T}_0(\vec{x_k})|d) \approx \mathcal{P}(\mathrm{T}_3(\vec{x_k})|d)$ is sufficiently small. These results are therefore consistent with expectations based on Gaussianity for the primordial large-scale structure in the Sloan volume.

\begin{table}\centering
\begin{tabular}{lcc}
\hline\hline
Structure type & $\mu_{\mathcal{P}(\mathrm{T}_i)}$ & $\sigma_{\mathcal{P}(\mathrm{T}_i)}$ \\
\hline
\multicolumn{3}{c}{Primordial large-scale structure ($a=10^{-3}$)} \\
Void & $0.07979$ & $5.4875 \times 10^{-5}$ \\
Sheet & $0.42022$ & $1.0240 \times 10^{-4}$ \\
Filament & $0.42022$ & $1.0412 \times 10^{-4}$ \\
Cluster & $0.07978$ & $5.6337 \times 10^{-5}$ \\
\hline\hline
\end{tabular}
\caption{Prior probabilities assigned by the \citet{Hahn2007a} web classifier to the different structures types, in the primordial large-scale structure ($a=10^{-3}$).}
\label{tb:prior_initial}
\end{table}

\begin{figure*}
\begin{center}
\includegraphics[width=0.49\textwidth]{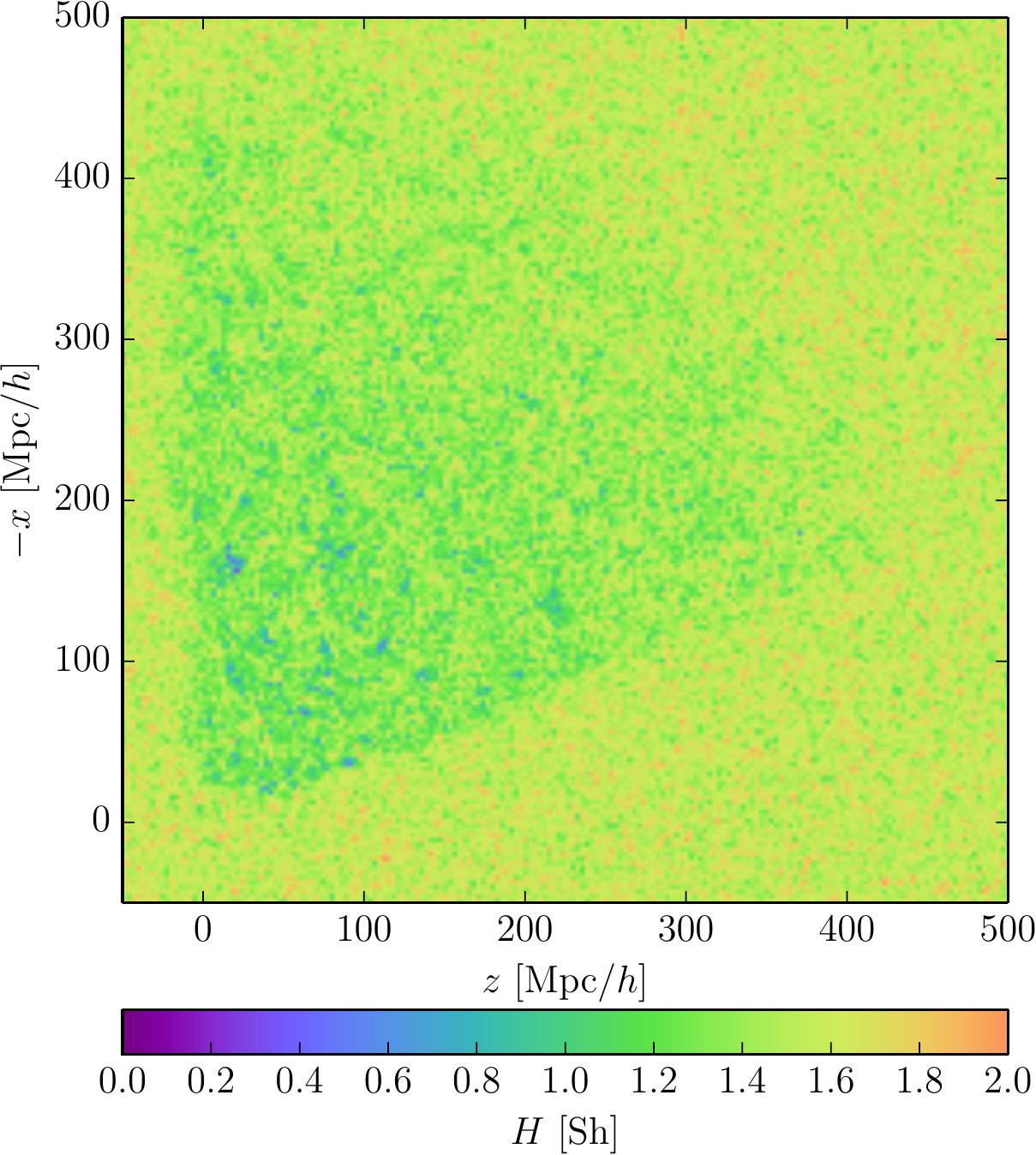}
\includegraphics[width=0.49\textwidth]{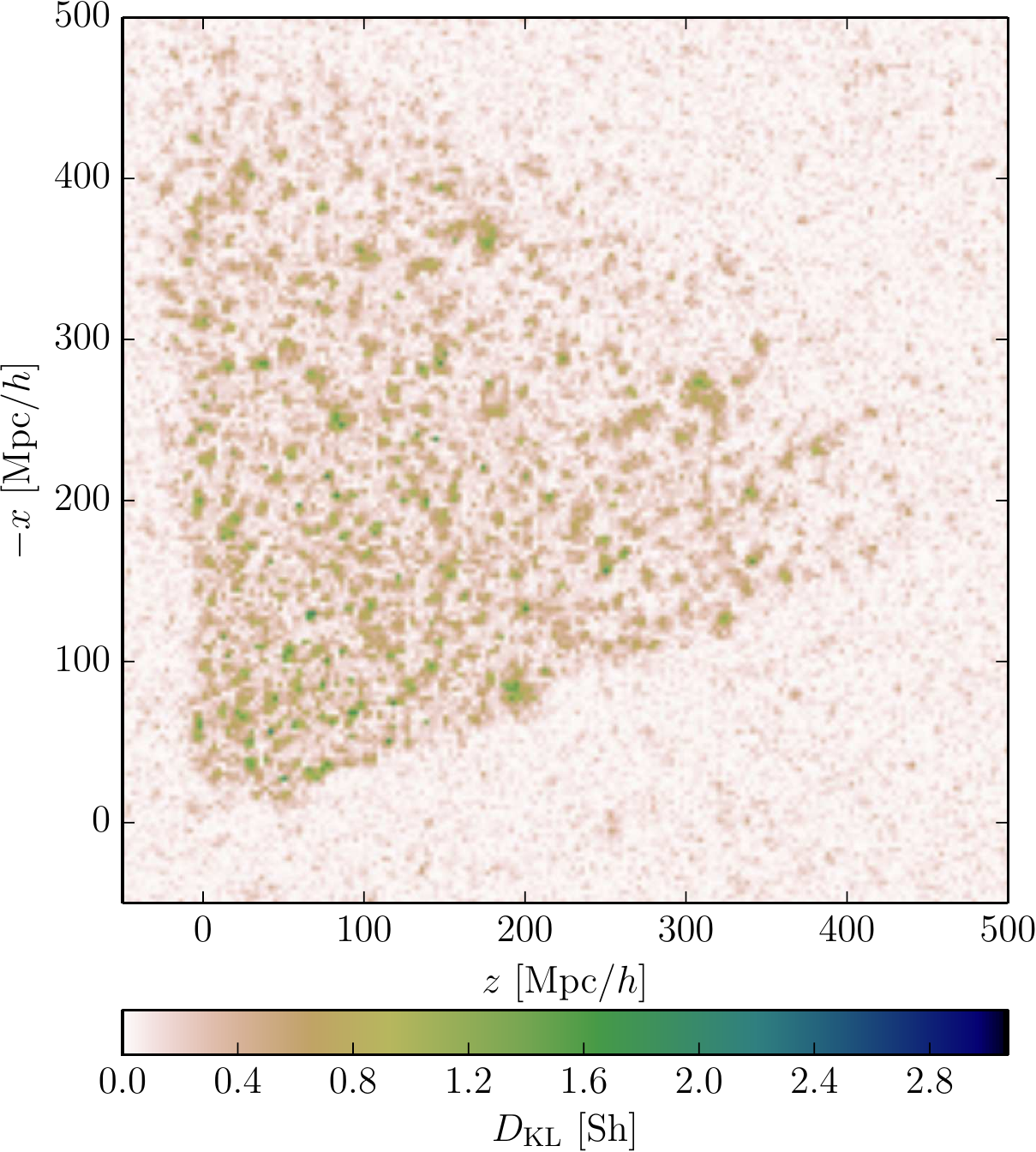}
\caption{Slices through the entropy of the structure types posterior (left panel) and the Kullback-Leibler divergence of the posterior from the prior (right panel), in the initial conditions. The entropy $H$, defined by equation \eqref{eq:definition_entropy}, quantifies the information content of the posterior pdf represented in figure \ref{fig:pdf_initial}, which results from fusing the information content of the prior and the data constraints. The Kullback-Leibler divergence $D_\mathrm{KL}$, defined by equation \eqref{eq:KL_divergence}, represents the information gained in moving from the prior to the posterior. It quantifies the information that has been learned on structure types by looking at SDSS galaxies.\label{fig:pdf_initial_entropy}}
\end{center}
\end{figure*}

In a similar fashion as in section \ref{sec:Probabilistic web-type cartography final}, the ensemble of samples permits us to propagate uncertainties to structure type classification and to characterize the strength of data constraints. In the left panel of figure \ref{fig:pdf_initial_entropy}, we show a slice through the voxel-wise entropy of the web-type posterior pdf in the initial conditions, defined by equation \eqref{eq:definition_entropy}. This function quantifies the information content of the posterior, which comes from both the prior and the data constraints. As in the final conditions, the entropy takes lower values inside the survey region. In the unobserved parts, the entropy fluctuates around 1.6~Sh, value which characterizes the information content of the prior. Using equation \eqref{eq:definition_entropy} (unconditional on the data) and the numbers given in table \ref{tb:prior_initial}, one can check that this number is consistent with the expectation. In the right panel of figure \ref{fig:pdf_initial_entropy}, we show a map of the Kullback-Leibler divergence of the posterior from the prior, which represents the information gain due to the data.

\subsection{Volume and mass filling fractions}
\label{sec:Volume and mass filling fractions initial}

We computed the volume and mass filling fractions (defined by equations \eqref{eq:definition_VFF} and \eqref{eq:definition_MFF}) of different structure types in the primordial large-scale structure in the Sloan volume. As for the final conditions, we kept only the regions where the survey response operator is strictly positive. Consequently, we obtained the posterior pdfs $\mathcal{P}(\mathrm{VFF}(\mathrm{T}_i)|d)$ and $\mathcal{P}(\mathrm{MFF}(\mathrm{T}_i)|d)$. Using a set of unconstrained Gaussian random fields, we also measured $\mathcal{P}(\mathrm{VFF}(\mathrm{T}_i))$ and $\mathcal{P}(\mathrm{MFF}(\mathrm{T}_i))$ and found that all these pdfs are well described by Gaussians, the means and standard deviations of which are given in table \ref{tb:initial_vff} and \ref{tb:initial_mff}.

All posterior quantities obtained are within two standard deviations of the corresponding prior means, and show smaller variance, as expected. Hence, all results obtained are consistent with Gaussian initial conditions.

\begin{table}\centering
\begin{tabular}{lcccc}
\hline\hline
Structure type & $\mu_{\mathrm{VFF}}$ & $\sigma_\mathrm{VFF}$ & $\mu_{\mathrm{VFF}}$ & $\sigma_\mathrm{VFF}$ \\
\hline
\multicolumn{1}{c}{} & \multicolumn{4}{c}{Primordial large-scale structure ($a=10^{-3}$)} \\
\multicolumn{1}{c}{} & \multicolumn{2}{c}{Posterior} & \multicolumn{2}{c}{Prior} \\
Void & $0.07994$ & $4.0221 \times 10^{-4}$ & $0.07977$ & $1.0200 \times 10^{-3}$ \\
Sheet & $0.41994$ & $6.1770 \times 10^{-4}$ & $0.42019$ & $1.7885 \times 10^{-3}$ \\
Filament & $0.42048$ & $6.3589 \times 10^{-4}$ & $0.42024$ & $1.7820 \times 10^{-3}$ \\
Cluster & $0.07964$ & $3.8043 \times 10^{-4}$ & $0.07980$ & $1.0260 \times 10^{-3}$ \\
\hline\hline
\end{tabular}
\caption{Mean and standard deviation of the prior and posterior pdfs for the volume filling fraction of different structure types in the primordial large-scale structure ($a=10^{-3}$).}
\label{tb:initial_vff}
\end{table}

\begin{table}\centering
\begin{tabular}{lcccc}
\hline\hline
Structure type & $\mu_{\mathrm{MFF}}$ & $\sigma_\mathrm{MFF}$ & $\mu_{\mathrm{MFF}}$ & $\sigma_\mathrm{MFF}$ \\
\hline
\multicolumn{1}{c}{} & \multicolumn{4}{c}{Primordial large-scale structure ($a=10^{-3}$)} \\
\multicolumn{1}{c}{} & \multicolumn{2}{c}{Posterior} & \multicolumn{2}{c}{Prior} \\
Void & $0.07958$ & $4.0122 \times 10^{-4}$ & $0.07941$ & $1.0163 \times 10^{-3}$ \\
Sheet & $0.41933$ & $6.1907 \times 10^{-4}$ & $0.41957$ & $1.7912 \times 10^{-3}$ \\
Filament & $0.42110$ & $6.3543 \times 10^{-4}$ & $0.42087$ & $1.7785 \times 10^{-3}$ \\
Cluster & $0.07999$ & $3.8206 \times 10^{-4}$ & $0.08015$ & $1.0293 \times 10^{-3}$ \\
\hline\hline
\end{tabular}
\caption{Mean and standard deviation of the prior and posterior pdfs for the mass filling fraction of different structure types in the primordial large-scale structure ($a=10^{-3}$).}
\label{tb:initial_mff}
\end{table}

\section{Evolution of the cosmic web}
\label{sec:Evolution of the cosmic web}

\begin{figure*}
\begin{center}
\includegraphics[width=0.49\textwidth]{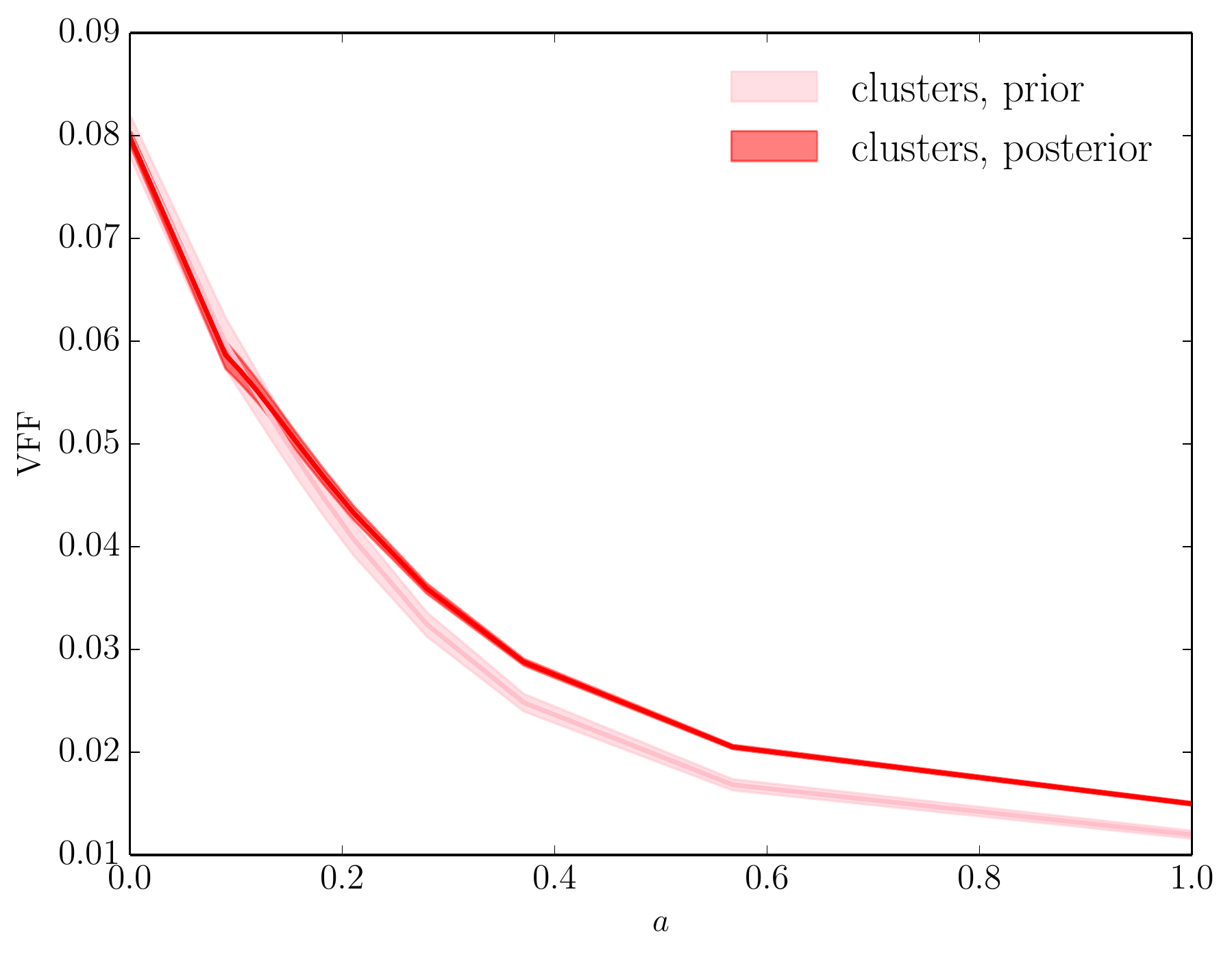}
\includegraphics[width=0.49\textwidth]{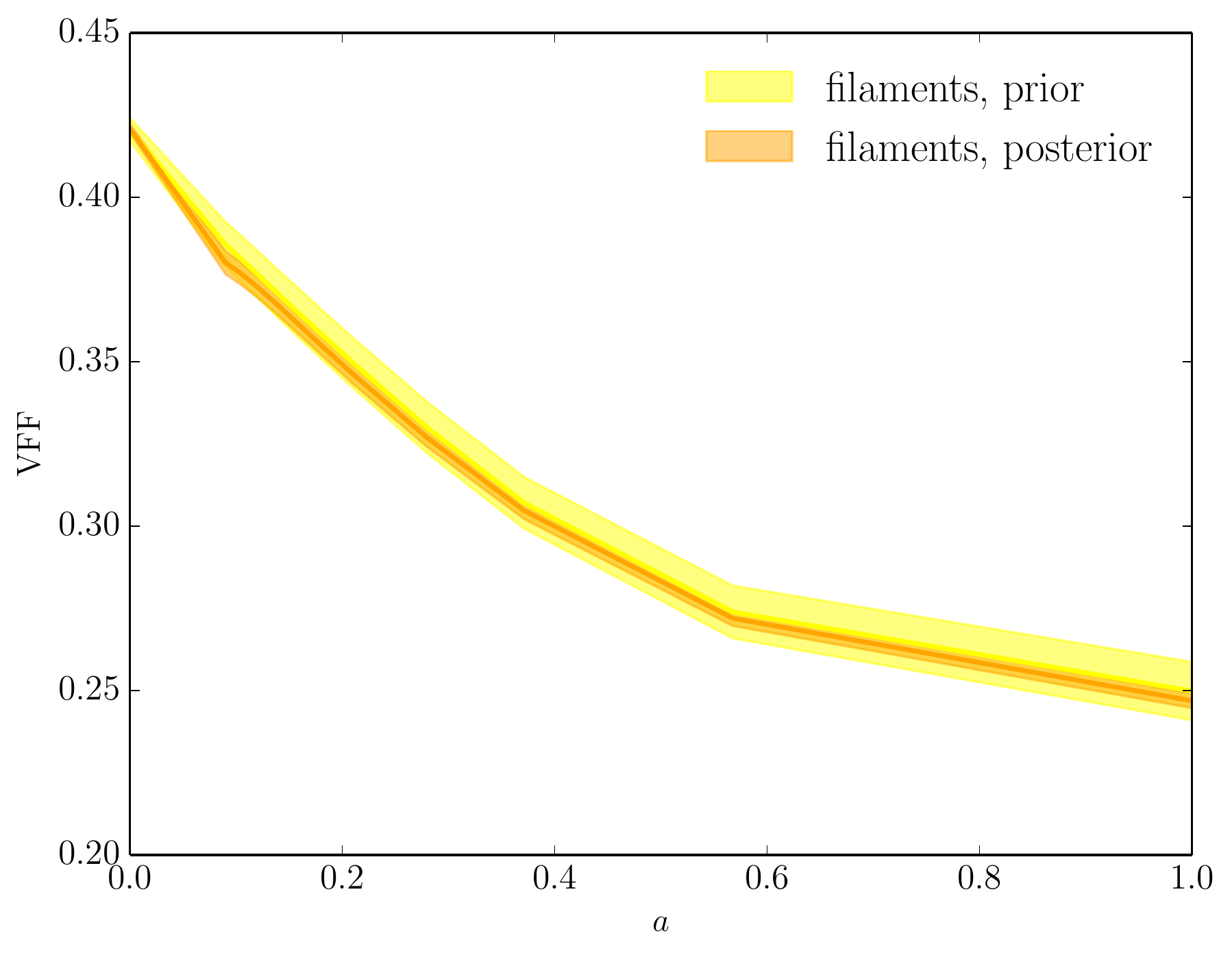} \\
\includegraphics[width=0.49\textwidth]{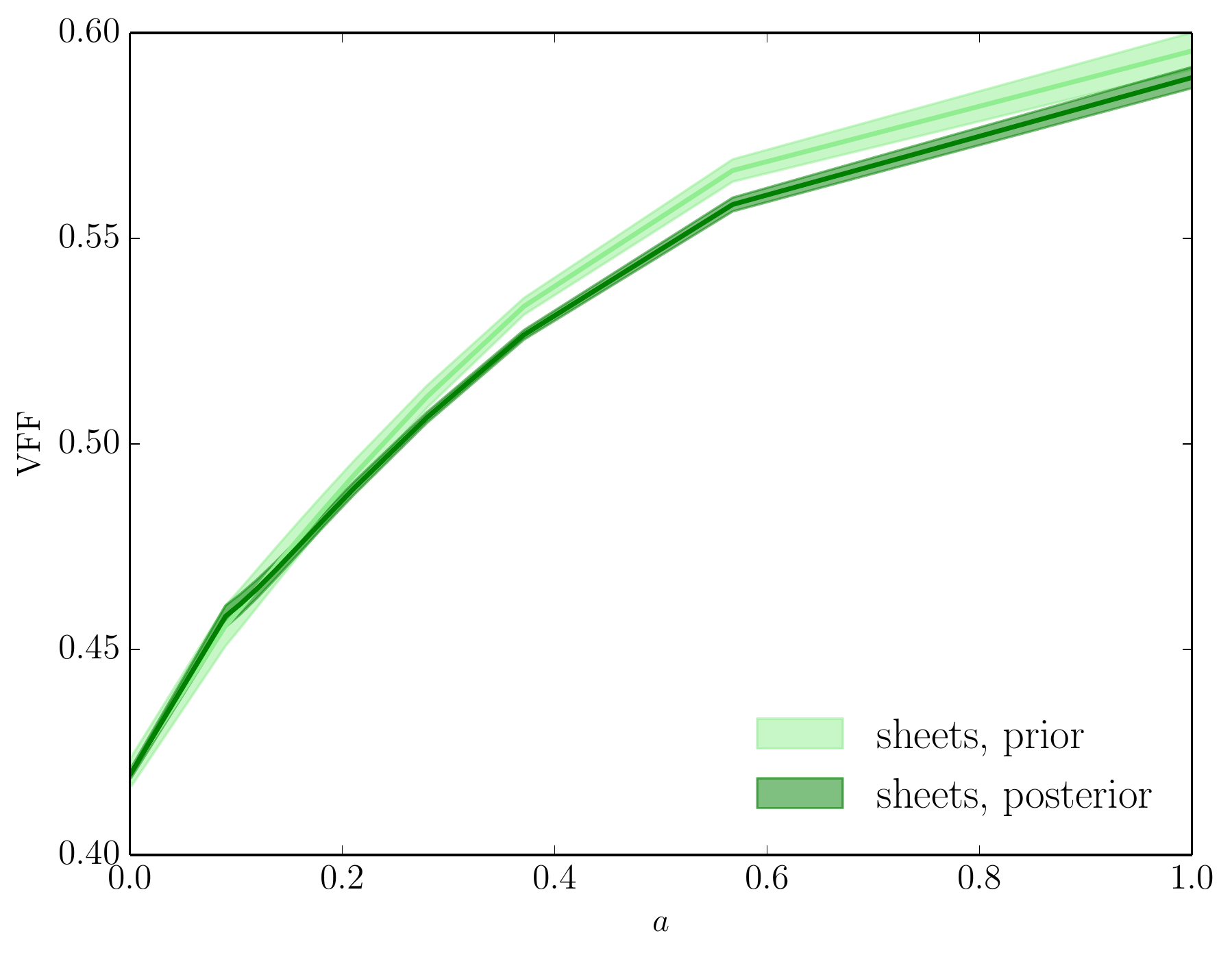}
\includegraphics[width=0.49\textwidth]{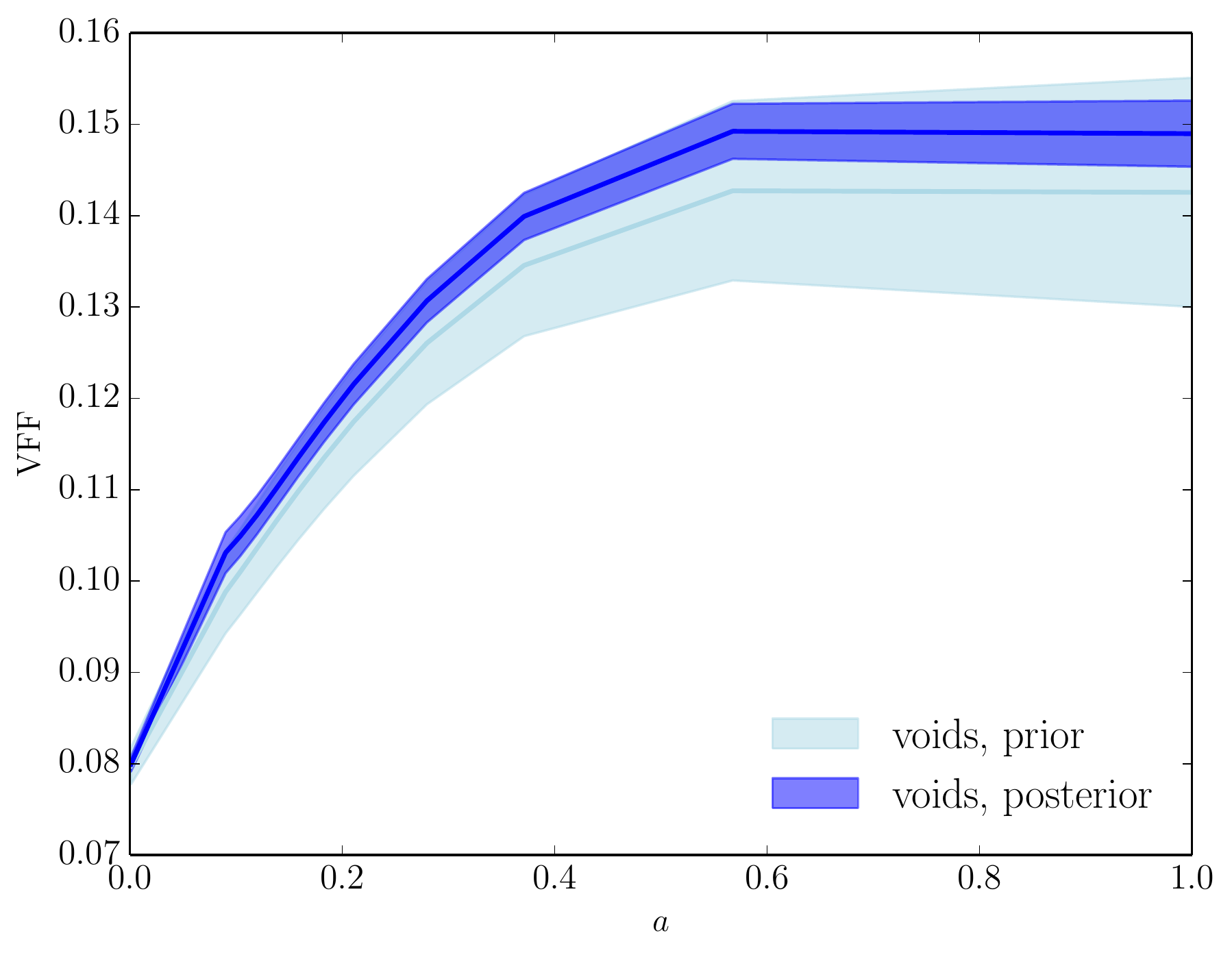} \\
\caption{Time evolution of the volume filling fractions of different structure types (from left to right and top to bottom: clusters, filaments, sheets, voids). The solid lines show the pdf means and the shaded regions are the 2-$\sigma$ credible intervals. Light colors are used for the priors and dark colors for the posteriors.\label{fig:evolution_vff}}
\end{center}
\end{figure*}

\begin{figure*}
\begin{center}
\includegraphics[width=0.49\textwidth]{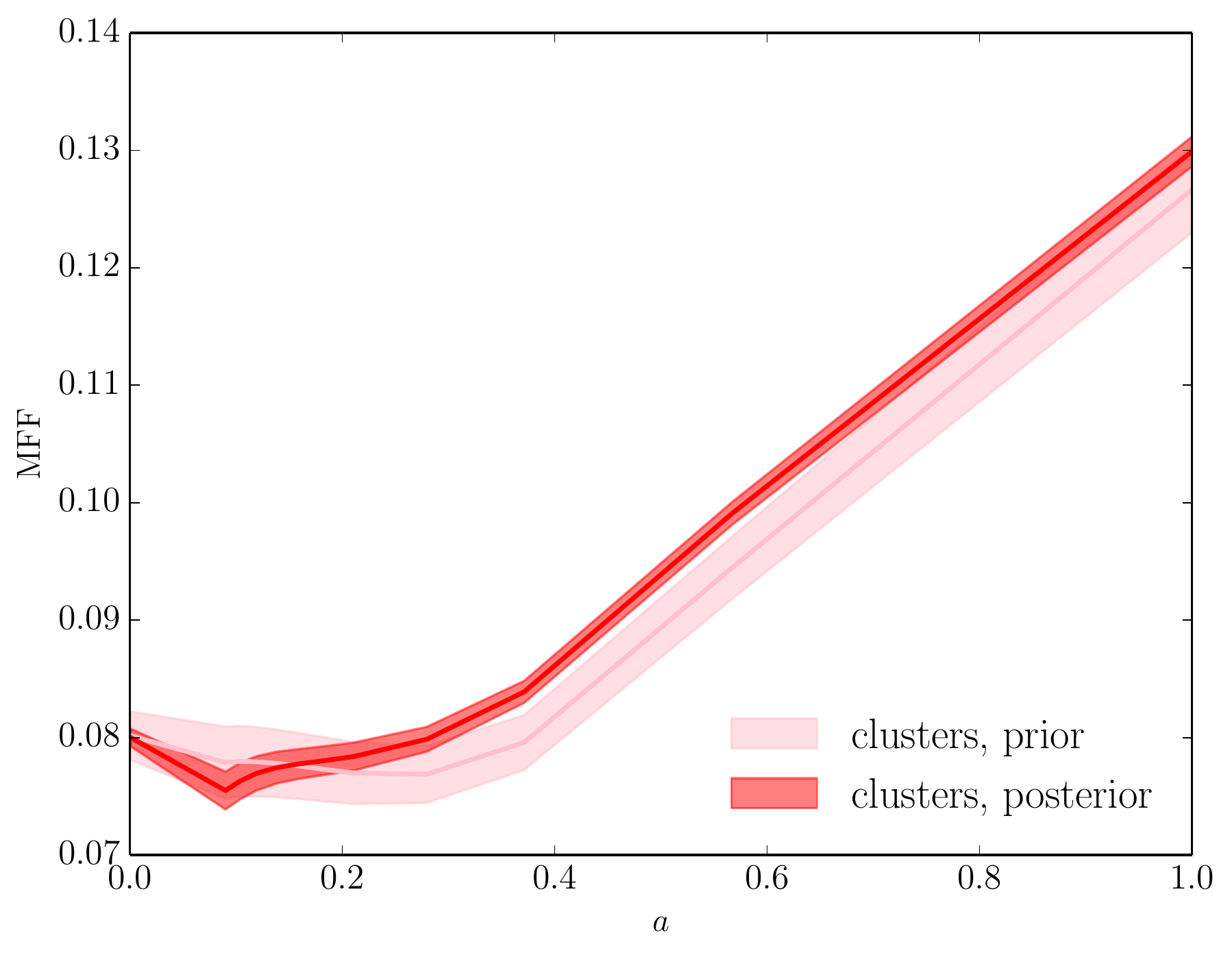}
\includegraphics[width=0.49\textwidth]{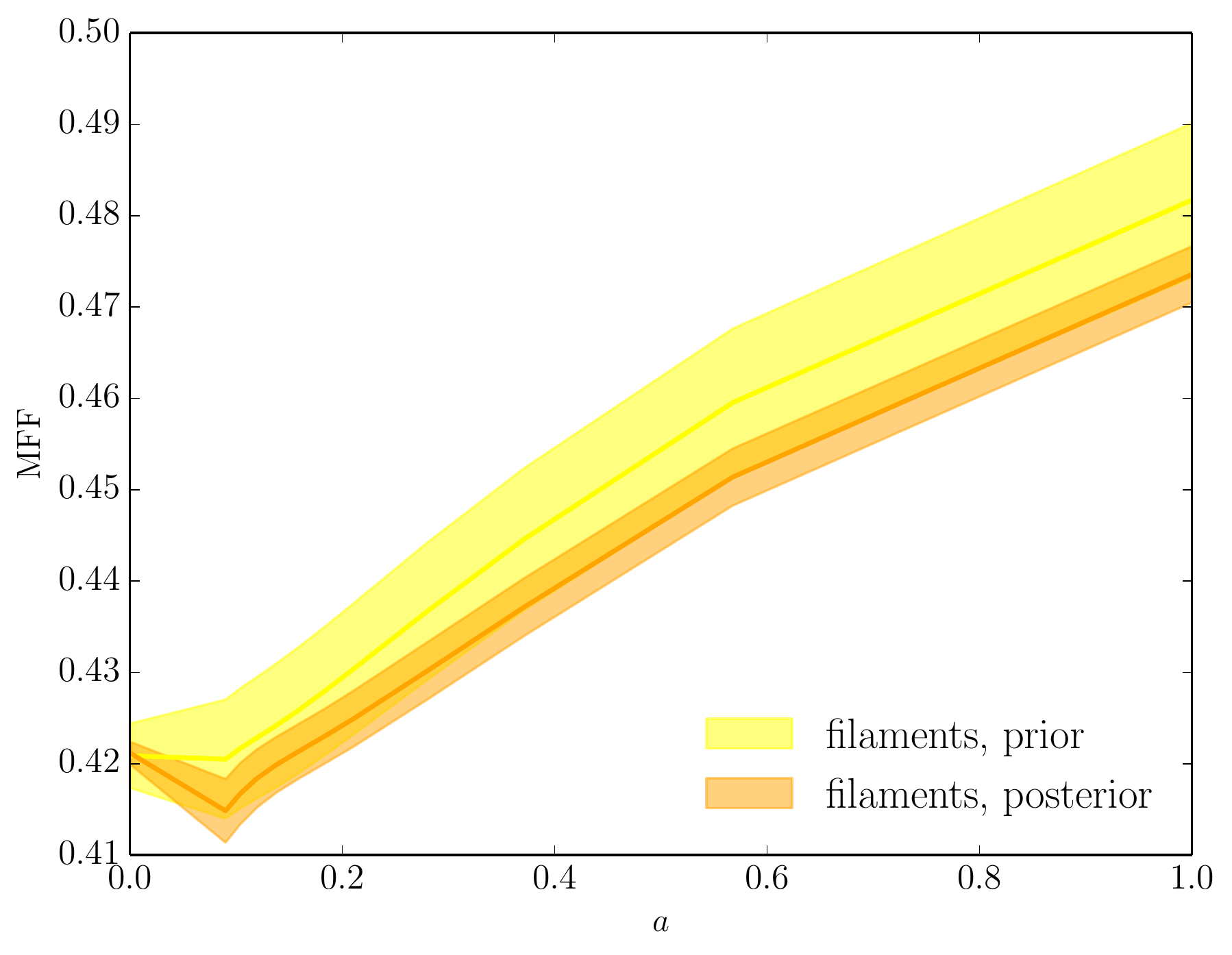} \\
\includegraphics[width=0.49\textwidth]{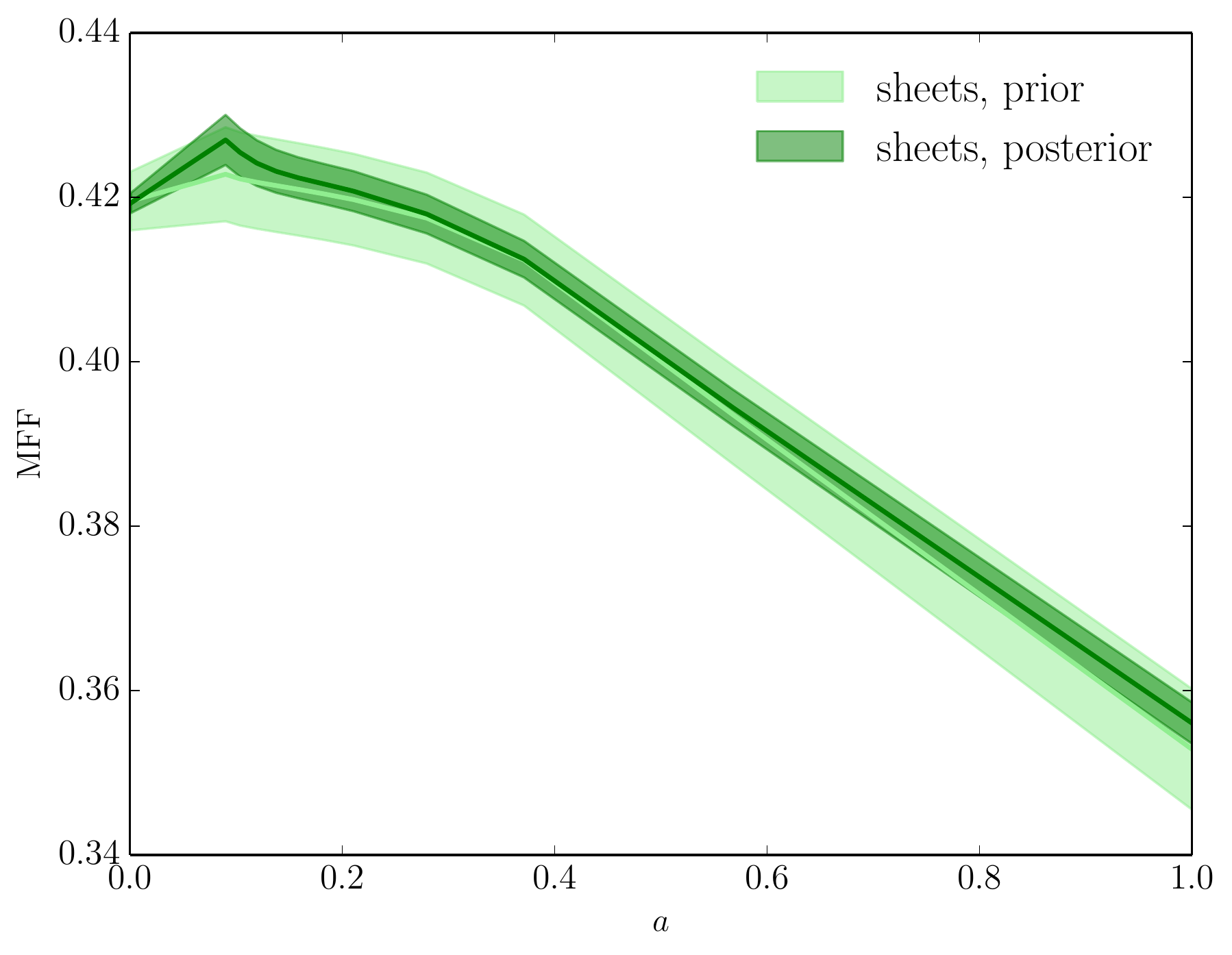}
\includegraphics[width=0.49\textwidth]{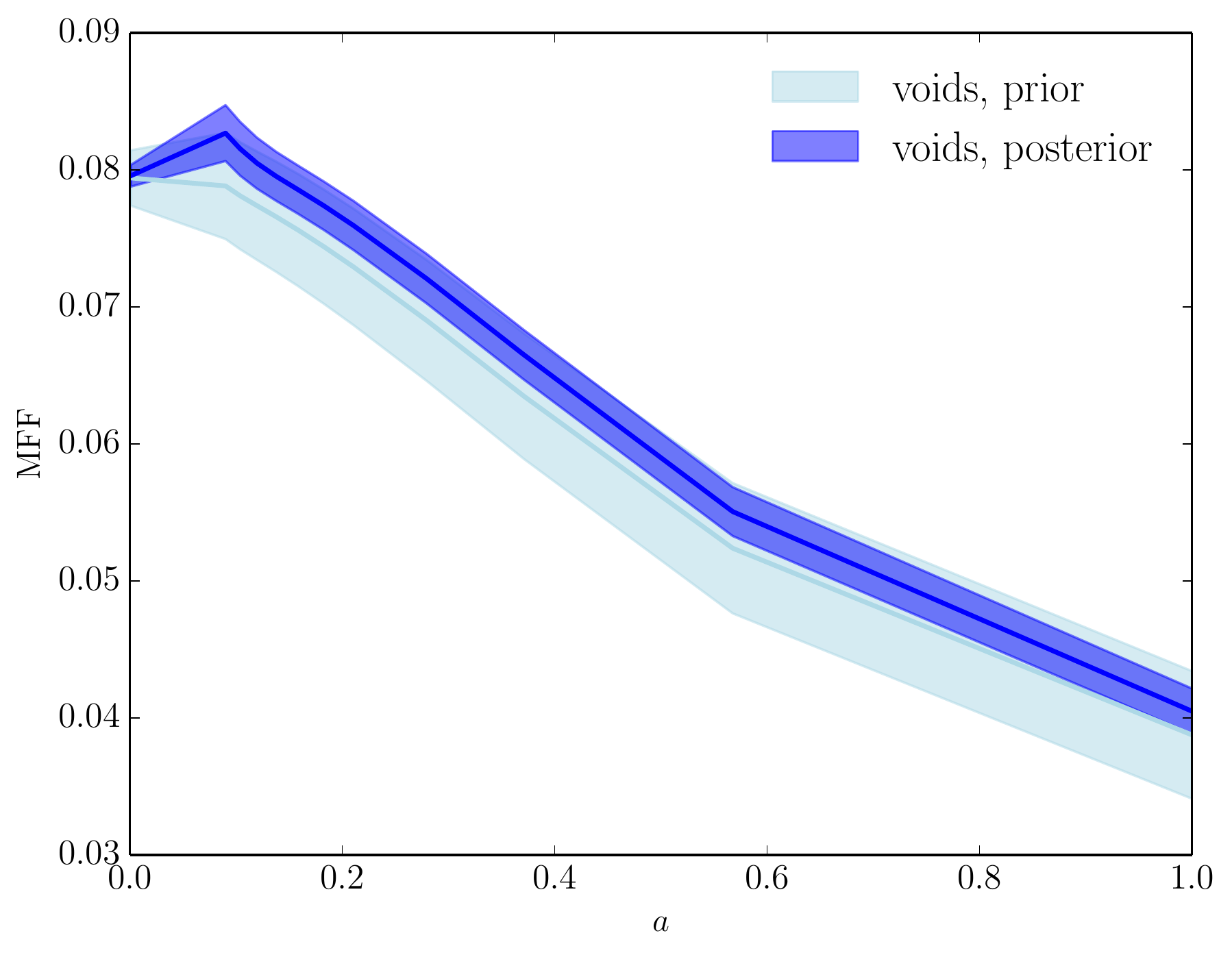} \\
\caption{Same as figure \ref{fig:evolution_vff} but for the mass filling fractions.\label{fig:evolution_mff}}
\end{center}
\end{figure*}

In addition to the inference of initial and final density fields, {\borg} allows to simultaneously analyze the formation history and morphology of the observed large-scale structure, a subject that we refer to as chrono-cosmography \citepalias{Jasche2015BORGSDSS}. In this section, we discuss the evolution of the cosmic web from its origin ($a=10^{-3}$, analyzed in section \ref{sec:The primordial large-scale structure}) to the present epoch ($a=1$, analyzed in section \ref{sec:The late-time large-scale structure}). To do so, we use $11$ snapshots saved during the \textsc{cola} filtering of our results (see section \ref{sec:Non-linear filtering of samples with COLA}). These are linearly separated in redshift from $z=10$ to $z=0$. We perform this analysis in the 1,097 samples filtered with \textsc{cola} considered in section \ref{sec:The late-time large-scale structure}. For each of these samples and for each redshift, we follow the procedure described in sections \ref{sec:Non-linear filtering of samples with COLA} and \ref{sec:Classification of the cosmic web} to compute the density field and to classify the structure types.

\subsection{Evolution of the probabilistic maps}

We followed the time evolution of the probabilistic web-type maps from the primordial (figure \ref{fig:pdf_initial}) to the late-time large-scale structure (figure \ref{fig:pdf_final}). In unconstrained regions, these maps show the evolution of the prior preference for specific structure types (see tables \ref{tb:prior_final} and \ref{tb:prior_initial}), in particular the breaking of the initial symmetry between voids and clusters and between sheets and filaments, discussed in section \ref{sec:Probabilistic web-type cartography initial}.

In data-constrained regions, the time evolution of web-type maps permits to visually check the expansion history of individual regions where the posterior probability of one specific structure is high. In particular, it is easy to see that, as expected from their dynamical definition, voids expand and clusters shrink in comoving coordinates, from $a=10^{-3}$ to $a=1$ (the reader is invited to compare the leftmost and rightmost panels of figures \ref{fig:pdf_final} and \ref{fig:pdf_initial}). Similarly, regions corresponding with high probability to sheets and filaments expand along two and one axis, respectively, and shrink along the others. This phenomenon is more difficult to see in slices, however, as the slicing plane intersects randomly the eigendirections of the tidal tensor. 

The time evolution of maps of the web-type posterior entropy (absolute and relative to the prior) also exhibit some interesting features. There, it is possible to simultaneously check the increase of the information content of the prior (from $H \approx 1.6$~Sh to $H \approx 1.4$~Sh) and the displacement of observational information operated by the physical model. As the large-scale structure forms in the Sloan volume, data constraints are propagated and the complex structure of the final entropy map (figure \ref{fig:pdf_final_entropy}), discussed in section \ref{sec:Probabilistic web-type cartography final}, takes shape.

\subsection{Volume filling fraction}

Our ensemble of snapshots allows us to check the time evolution of global characterizations of the large-scale structure such as the volume and mass filling fractions of different structures. As in sections \ref{sec:Volume and mass filling fractions final} and \ref{sec:Volume and mass filling fractions initial}, we computed these quantities using only the volume where the survey response operator is non-zero. In figure \ref{fig:evolution_vff}, we plot these VFF as a function of the scale factor. There, the solid lines correspond to the pdf means and the shaded regions to the 2-$\sigma$ credible intervals, with light colors for the priors and dark colors for the posteriors.

The time variation of the VFF in figure \ref{fig:evolution_vff} is consistent with the expected dynamical behavior of structures. As voids and sheets expand along three and two axes, respectively, their volume fraction increases. Here, the posterior probabilities are mild updates of this prediction. Conversely, as clusters and filaments shrink along three and two axes, respectively, their volume fraction decreases. An explanation for the substantial displacement of the posterior from the prior, observed for clusters, can be found in section \ref{sec:Volume and mass filling fractions final}.

As already noted, the VFF is a very sensitive function of the precise definition of structures, grid size, density assignment scheme, smoothing scale, etc. For this reason, even for prior probabilities, our results can be in qualitative disagreement with previous authors \citep[e.g. figure 23 in][]{Cautun2014}, due to their very different definitions of structures. Therefore, we only found relevant to compare our posterior results with the prior predictions based on unconstrained realizations. The same remark applies to the MFF in the following section.

\subsection{Mass filling fraction}

In figure \ref{fig:evolution_mff}, we show the time evolution of the mass filling fractions using the same plotting conventions. Results are consistent with an interpretation based on large scale flows of matter. According to this picture, voids always loose mass while clusters always become more massive. The behavior of sheets and filaments can in principle be more complex, since these regions have both inflows and outflows of matter depending on the detail of their expansion profiles. In our setup, we found that the number of axes along which there is expansion dominates in the determination of the balance of inflow versus outflow, for global quantities such as the MFF. Therefore, filaments always gain mass and sheets always loose mass. Summing up our prior predictions, as they expand along at least two axes, matter flows out of voids and sheets and streams towards filaments and clusters.

The posterior probabilities slightly update this picture. Observations support smaller outflowing of matter from voids. For structures globally gaining matter, the priors are displaced towards less massive filaments and more massive clusters. All posterior predictions fall within the $\sim2$-$\sigma$ credible interval from corresponding prior means.

\section{Summary and Conclusion}
\label{sec:Conclusion}

Along with \citet{Leclercq2015DMVOIDS}, this work exploits the high quality of inference results produced by the application of the Bayesian code {\borg} \citep{Jasche2013BORG} to the Sloan Digital Sky Survey main galaxy sample \citepalias{Jasche2015BORGSDSS}. We presented a Bayesian cosmic web analysis of the nearby Universe probed by the northern cap of the SDSS and its surrounding. In doing so, we produced the first probabilistic, four-dimensional maps of dynamic structure types in real observations.

As described in section \ref{sec:Bayesian large-scale structure inference with BORG}, our method relies on the physical inference of the initial density field in the LSS (\citealp{Jasche2013BORG}; \citetalias{Jasche2015BORGSDSS}). Starting from these, we generated a large set of data-constrained realizations using the fast {\cola} method (section \ref{sec:Non-linear filtering of samples with COLA}). The use of 2LPT as a physical model in the inference process and of the fully non-linear gravitational dynamics, provided by {\cola}, as a filter allowed us to describe structures at the required statistical accuracy, by very well representing the full hierarchy of correlation functions. Even though initial conditions were inferred with the approximate 2LPT model, we checked that the clustering statistics of constrained non-linear model evaluations agree with theoretical expectations up to scales considered in this work. As described in section \ref{sec:Classification of the cosmic web}, we used the dynamic web-type classification algorithm proposed by \citet{Hahn2007a} to dissect the cosmic web into voids, sheets, filaments, and clusters.

In sections \ref{sec:The late-time large-scale structure} and \ref{sec:The primordial large-scale structure}, we presented the resulting maps of structures in the final and initial conditions, respectively, and studied the distribution of global quantities such as volume and mass filling fractions. In section \ref{sec:Evolution of the cosmic web}, we further analyzed the time evolution of our results, in a rigorous chrono-cosmographic framework.

For all results presented in this paper, we demonstrated a thorough capability of uncertainty quantification. Specifically, for all inferred maps and derived quantities, we got a probabilistic answer in terms of a prior and a posterior distribution. The variation between samples of the posterior distribution quantifies the remaining uncertainties of various origins (in particular noise, selection effects, survey geometry and galaxy bias, see \citetalias{Jasche2015BORGSDSS} for a detailed discussion). Building upon our accurate probabilistic treatment, we looked at the entropy of the structure type posterior and at the relative entropy between posterior and prior. In doing so, we quantified the information gain due to SDSS galaxy data with respect to the underlying dynamic cosmic web and analyzed how this information is propagated during cosmic history. This study constitutes the first link between cosmology and information theory using real data.

In summary, our methodology yields an accurate cosmographic description of web types in the non-linear regime of structure formation, permits to analyze their time evolution and allows a precise uncertainty quantification in a full-scale Bayesian framework. These inference results can be used for a rich variety of applications, ranging from studying galaxies inside their environment to cross-correlating with other cosmological probes. They count among the first steps towards accurate chrono-cosmography, the subject of simultaneously analyzing the morphology and formation history of the inhomogeneous Universe.

\textit{Note added:} As we were finalizing this paper for submission, the works by \citet{Zhao2015} and \citet{Shi2015} appeared where the relationship between halos and the cosmic web environment defined by the tidal tensor is being studied.

\acknowledgments

We thank Jacopo Chevallard, Guilhem Lavaux, Joseph Silk and Mat\'ias Zaldarriaga for stimulating discussions. Special thanks go to St\'ephane Rouberol for his support during the course of this work, in particular for guaranteeing flawless use of all required computational resources. The results described in this work have been presented at the International Astronomical Union Symposia 306 (``Statistical Challenges in 21st Century Cosmology'') and 308 (``The Zel'dovich Universe''). We would like to express our gratitude to the organizers. FL acknowledges funding from an AMX grant (\'Ecole polytechnique ParisTech). JJ is partially supported by a Feodor Lynen Fellowship by the Alexander von Humboldt foundation. BW acknowledges funding from an ANR Chaire d'Excellence (ANR-10-CEXC-004-01) and the UPMC Chaire Internationale in Theoretical Cosmology. This work has been done within the Labex \href{http://ilp.upmc.fr/}{Institut Lagrange de Paris} (reference ANR-10-LABX-63) part of the Idex SUPER, and received financial state aid managed by the Agence Nationale de la Recherche, as part of the programme Investissements d'avenir under the reference ANR-11-IDEX-0004-02. This research was supported by the DFG cluster of excellence ``\href{www.universe-cluster.de}{Origin and Structure of the Universe}''.

\bibliography{structuretypes}

\end{document}